\documentclass[a4paper,12pt]{article}
\usepackage{amsmath,amsfonts,amssymb}
\usepackage[mathscr]{eucal}
\usepackage{natbib}
\newtheorem{lemma}{Lemma}
\newtheorem{definition}[lemma]{Definition}
\newtheorem{theorem}[lemma]{Theorem}\newtheorem{conjecture}[lemma]{Conjecture}

\newtheorem{corollary}[lemma]{Corollary}

\newtheorem{condition}[lemma]{Condition}
\newtheorem{postulate}[lemma]{Postulate}
\newtheorem{fact}[lemma]{Fact}

\newcommand{\qed}{\bf Q.E.D.} 
\newenvironment{proof}{{\bf Proof}: }{\;\;\qed}

\newcommand{\reals}{{\mathbb R}}
\newcommand{\naturals}{{\mathbb N}}

\begin{document}
\title{On Infinite EPR-like Correlations}
\author{%
Leszek Wro{\' n}ski  and Tomasz Placek%
\thanks{We read earlier versions of this paper at the seminar `Chaos and Quantum Information' held at the Jagiellonian University in Krak{\' o}w on April 16, 2007 and at the seminar `On Determinism' held at the University of Bonn on April 20, 2007. For comments and valuable discussions we are grateful to the audiences, and in particular, to Dr. Thomas M{\" uller}. Authors' address: Department of Philosophy, Jagiellonian University,
        Grodzka 52, PL 31-044 Krak{\' o}w, Poland;\;
LW's email: {\tt elwro1@gmail.com} and
TP's email:  {\tt uzplacek@cyf-kr.edu.pl}.%
}}
\date{\today}
\maketitle
\begin{abstract}
  The paper investigates, in the framework of branching space-times,
  whether an infinite EPR-like correlation which does not involve
  finite EPR-like correlations is possible.
\end{abstract}

\section{Introduction} The question addressed in the present paper is
the following: Is an infinite EPR-like correlation which does not
involve finite EPR-like correlations possible?  To explain what finite
/ infinite refers to in this context, recall the essence of the
Einstein-Podolski-Rosen's (\citeyear{einstein35}) set-up: two
particles emerging from a source are subjected to position (or
momentum) measurements at distant locations in such a way that the two
measurement events are space-like separated.  The result is a
(perfect) correlation between outcomes of position measurements (or
momentum measurements): if the position measured on the first particle
is such-and-such, then the position measured on the second particle is
certainly such-and-such. The correlation is finite, as it refers to
outcomes of two measurements, performed, say, at the left and at the
right station. In the subsequent investigations of EPR-like
correlations (and of Bell's theorem), set-ups with more measurement
stations were considered.  Thus, \citet{mermin90} considers correlated
triples of outcomes of spin measurements -- in this case a trio of
outgoing particles is subjected to spin measurements at three
stations.  And \citet{greenberger89} investigate a set-up with four
particles and hence: four measurement stations and quadruples of
correlated outcomes. Accordingly, in our usage, the size of a
correlation (and hence the distinction: finite vs.  infinite) refers
to the number of correlated measurement outcomes.

Clearly, the crux of our problem is the notion of possibility, as it
occurs in our question. It is perhaps interesting to learn that an
infinite EPR-like correlation which does not involve finite EPR-like
correlations is logically / mathematically possible; yet, we are after
a sharper notion of possibility. On the other extreme, our
investigations have nothing to do with experimental possibility, i.e. the
possibility of producing an experimental set-up for infinite
correlations. Somewhat similarly, we leave aside the question whether
quantum theory permits infinite EPR-like correlations.  Instead we
focus on what we take to be essential aspects of EPR-like
correlations: its spatiotemporal and modal features.  As for the
former, we assume that the experiment occurs in Minkowski space-time.
The modal aspect is seen in the popular diagnosis of EPR, which says
that although every outcome of a single measurement is possible,
certain combinations of outcomes (or, equivalently, a joint outcome)
is impossible. To illustrate, although $+$ as well as $-$ are possible
outcomes of spin measurement, as performed on one particle, the
outcome $++$, that is $+$ on both the particles, is impossible. A
final word of warning: we neglect the probabilistic aspect of EPR-like
correlations, and we do it for two reasons. First, we focus on perfect
correlations and anti-correlations, and we read the extreme
probabilities as `it must happen / it cannot happen'. Second, we
believe that in EPR correlations the modal aspect has a conceptual
priority over the probabilistic one.

Given the above assumptions, the essence of EPR-like correlations is
as follows: in Minkowski space-time, there is a number (finite or not)
of measurement events, every two of which are space-like separated.
For each measurement event there is a set of possible `single'
outcomes. Yet, certain combinations of single outcomes are impossible.

EPR-like correlations, as described above, can be rigorously
investigated in (a non-probabilistic version of) \emph{branching
  space-times} (BST), a theory proposed by
\citet{belnap92}.\footnote{For the updated version of this paper,
  see its ``postprint''.} The BST framework
rigorously combines modality and (rudiments of) special relativity. It
has been used to diagnose Bell's theorem.\footnote{Cf.
  \citet{belnap-szabo} and \citet{placek00}.} In this theory, a
feature analogous to pre-probabilistic EPR-like correlation is called
''modal funny business'', which is defined so as to capture the idea
that a certain combination of otherwise possible outcomes of
space-like separated measurement events is impossible. The
investigation of this notion brought \citet{mueller05} to ask
whether infinite modal funny business which would not involve finite
modal funny business was possible. It is this question that sparked
our interest in the issues discussed here. In a sense, the question
was answered in the positive by M{\"u}ller, Belnap, and Kohei
(\citeyear{muller06:_funny}).  They produced a set-theoretical structure
(called M2 and described here in Section~\ref{sec3}) which satisfies
all the axioms of BST and exhibits infinite modal funny business
without there being finite modal funny business.  Yet, the structure
has no relation to any space-time, incl.  Minkowski space-time. Thus,
the intriguing question remains: namely, can there be in Minkowski
space-time a case of infinite modal funny business which does not
involve finite modal funny business?

This state of affairs brings in our first task. Contrary to first
appearances, BST has models the possible histories (analogous to
Lewis-style possible worlds) of which are not associated with
Minkowski space-time (or even any space-time). Thus, to investigate
whether infinite EPR-like correlations are possible, we had better
single out those BST models in which histories are isomorphic to
Minkowski space-times.  We call such models Minkowskian branching
structures (MBS for short). In defining this notion we follow the lead
of \citet{muller_nato}, yet with two crucial diversions. First, we
remove M{\"u}ller's finiteness assumptions, as they prohibit
introduction of `interesting' infinite structures. Second, we improve
on M{\"u}ller's failed proof of the most desired feature of MBS,
namely that every history is isomorphic to Minkowski space-time. To
this end we assume a certain topological postulate.

As our second task, we single out two postulates such that each
generates infinite funny business in an arbitrary BST model, and such
that if none holds in a BST model, the model is free from infinite
funny business.

We finally show that in an MBS, if there is infinite funny business, a
set in $\Re^4$ which generates it must be very strangely located. We
also show that truth of the first postulate (A) requires a converging
sequence of measurement events. We finally exhibit an MBS model in
which the other postulate~(B) is true, yet the model has an odd
feature, which (we conjecture) is necessary for the truth of the
postulate. Thus, our findings strongly say against a possibility of
INFFB in physically motivated models of BST.

The paper is organized as follows. In Section~\ref{sec1} we review
some definitions and facts of BST which we need later.
Section~\ref{sec2} defines and discusses Minkowskian branching
structures. Section~\ref{sec3} provides definitions of modal funny
business for general BST, introduces two postulates, and relates the
occurrence / non-occurrence of infinite funny business in BST to the
satisfaction (or not) of those postulates.  Section~\ref{sec5} links
the above to Minkowskian branching structures: it asks what the
postulates presuppose of MBS, and hence, for what price one can have
in Minkowski space-time an infinite EPR-like correlation which does
not involve finite correlations.  The final section~\ref{concl} states
our conclusions and poses some open problems.

\section{Branching Space-Times}{\label{BST}}\label{sec1}
The theory of Branching Space-Times (BST), as presented by Nuel Belnap
in 1992 \cite{belnap92}, combines objective indeterminism and
relativity in a rigorous way. Its primitives are a nonempty set $W$
(called ``Our World'', interpreted as the set of all possible point
events) and a partial ordering $\leqslant$ on $W$, interpreted as a
``causal order'' between point events.

There are no ``Possible Worlds'' in this theory; there is only one
world, Our World, containing all that is (timelessly) possible.
Instead, a notion of ``history'' is used, as defined below:

\begin{definition}{\label{updir}} A set $h \subseteq W$ is upward-directed iff
$\forall e_1, e_2 \in h \; \exists e \in h$ such that $e_1 \leqslant e$ and $e_2 \leqslant e$.

A set $h$ is maximal with respect to the above property iff $\forall g \in W$ such that $g \varsupsetneq h$, $g$
is not upward-directed.

A subset $h$ of $W$ is a history iff it is a maximal upward-directed set.

For histories $h_1$ and $h_2$, any maximal element in $h_1 \cap h_2$
is called a choice point for $h_1$ and $h_2$.
\end{definition}

A very important feature of BST is that histories are closed downward:
if $e_1 \leqslant e_2$ and $e_1 \notin h$, then $e_2 \notin h$. In
other words, there is no backward branching among histories in BST. No
two incompatible events are in the past of any event; equivalently:
the past of any event is ``fixed'', containing only compatible events.

We will now give the definition of a BST model; for more information
about BST in general see \cite{belnap92}.

\begin{definition}{\label{BSTmodel}}
$\langle W, \leqslant \rangle$ where $W$ is a nonempty set and $\leqslant$ is a partial ordering on $W$ is a
model of BST if and only if it meets the following requirements:
\begin{enumerate}
\item The ordering $\leqslant$ is dense.
\item $\leqslant$ has no maximal elements.
\item Every lower bounded chain in $W$ has an infimum in $W$.
\item Every upper bounded chain in $W$ has a supremum in every
history that contains it.
\item (Prior choice principle ('PCP')) For any lower bounded chain $O \in
h_1 - h_2$ there exists a point $e \in W$ such that $e$ is maximal in $h_1 \cap h_2$ and $\forall e^{'} \in O \;
e < e^{'}$.
\end{enumerate}

\end{definition}

\section{Introducing Minkowskian Branching Structures}{\label{MBS}}\label{sec2}
BST is a frugal theory, since it allows for models whose histories
hardly have spacetimes, and the only 'spatiotemporal' notion is that
of the ordering $\leqslant$. We are going to single out a class of BST
models, in which histories occur in spacetimes, and moreover, all
these spacetimes are Minkowskian. This part of our work is based on
M\"{u}ller's [2002] theory.

The points of the Minkowskian space-time are elements of $\reals^4$, e.g. $x = \langle x^0, x^1, x^2, x^3
\rangle$, where the first element of the quadruple is the time coordinate. The Minkowskian space-time distance (Lorentz interval)
is a function $D^2_M : \reals^4 \times \reals^4 \rightarrow \reals$ defined as follows (for $x,y \in \reals^4$):
\begin{equation}
D^2_M(x,y) := -(x^0 - y^0)^2 + \sum_{i=1}^3(x^i - y^i)^2
\end{equation}
The natural ordering on the Minkowski space-time, call it ``Minkowskian ordering $\leqslant_M$'', is defined as
follows ($x,y \in \reals^4$):
\begin{equation}
x \leqslant_M y \mbox{ iff } D^2_M(x,y) \leqslant 0 \mbox{ and } x^0 \leqslant y^0
\end{equation}

We will say that two points $x, y \in \reals^4$ are space-like related
(``SLR'' for short) iff neither $x \leqslant_M y$ nor $y \leqslant_M
x$. Naturally, $x <_M y$ iff $x \neq y$ and $x \leqslant_M y$.

Now we need to provide a framework for ``different ways in which
things can happen'' and for filling the space-times with content. For
the first task we will need a set $\Sigma$ of labels $\sigma, \eta,
...$. (In contrast to \cite{muller_nato}, we allow for any cardinality
of $\Sigma$). For the second task, we will use a so called ``state''
function $S : \Sigma \times \reals^4 \rightarrow P$, where $P$ is a
set of point properties (on this we just quote M\"{u}ller saying
``finding out what the right $P$ is is a question of physics, not one
of conceptual analysis'').

One could ask about the reasons for an extra notion of a ``scenario''.
Why don't we build histories just out of points from $\reals^4 \times
P$? The reason is that a member of BST's Our World has a fixed past.
If two different trains of events lead to exactly the same event $E
\in \reals^4 \times P$, the situation gives rise to two different
point events, two different members of $W$. In contrast: for a point
$\langle x, p_0 \rangle$ from $\reals^4 \times P$ there can exist two
different points $\langle y, p_1 \rangle$ and $\langle y, p_2 \rangle$
from $\reals^4 \times P$ such that $y <_M x$. This would be a case of
backward branching, so the set $\reals^4 \times P$ is not a good
candidate for the master set $W$ of any BST model.

The idea behind the concept of scenario is that every scenario
corresponds to a $\reals^4$ space filled with content\footnote{Fix a
  scenario $\alpha$. The above mentioned corresponding space filled
  with content is $A \subseteq \reals^4 \times P$ such that $\langle
  x, p \rangle \in A$ iff $S(\alpha,x)=p$.}, where the content derives
from the elements of $P$. Assuming a certain state function $S$ is
given, for any $\sigma, \eta \in \Sigma$ the set $C_{\sigma\eta}
\subset \reals^4$ is the set of ``splitting points'' between scenarios
$\sigma$ and $\eta$, intuitively: the set of points in which a choice
between the two scenarios is made. All members of $C_{\sigma\eta}$
have to be space-like related. Of course a choice between $\sigma$ and
$\eta$ is a choice between $\eta$ and $\sigma$, so $C_{\sigma\eta} =
C_{\eta\sigma}$. The BST axiom of prior choice principle motivates our
postulate that any two different scenarios split. Formally: $\forall
\sigma, \eta \in \Sigma \; (\sigma \neq \eta \Rightarrow
C_{\sigma\eta} \neq \emptyset)$.

The next requirement concerns triples of scenarios. Any set
$C_{\sigma\eta}$ determines a region in which both scenarios coincide:
namely, that part of $\reals^4$ that is not in the Minkowskian sense
strictly above any point from $C_{\sigma\eta}$. Following M\"{u}ller
we call it the region of overlap $R_{\sigma\eta}$ between scenarios
$\sigma, \eta$ defined as below:
\begin{equation}\label{overlap}
R_{\sigma\eta} := \{ x \in \reals^4 | \neg \exists y \in C_{\sigma\eta} \; y <_M x \}
\end{equation}
(Of course it follows that for any $\sigma, \eta \in \Sigma
C_{\sigma\eta} \subseteq R_{\sigma\eta}$.) Assuming the sets
$C_{\sigma\eta}$ and $C_{\eta\gamma}$ are given, we get two regions of
overlaps $R_{\sigma\eta}$ and $R_{\eta\gamma}$. At the points in the
intersection of those two regions $\sigma$ coincides with $\eta$ and
$\eta$ coincides with $\gamma$, therefore by transitivity $\sigma$
coincides with $\gamma$. In general we can say that for any $\sigma,
\eta, \gamma \in \reals^4$
\begin{equation}\label{overlap2}
R_{\sigma\gamma} \supseteq R_{\sigma\eta} \cap R_{\eta\gamma}
\end{equation}
which translated to a requirement on sets of splitting points is
\begin{equation}\label{overlap3}
\forall x \in C_{\sigma\gamma} \exists y \in C_{\sigma\eta} \cup C_{\eta\gamma} y \leqslant_M x.
\end{equation}

In his paper M\"{u}ller put another requirement on $C_{\sigma\eta}$:
finitude. The motivation was to exclude splitting along a
``simultaneity slice''. The strong requirement of finitude excludes
however many more types of situations, in which splitting is not
continuous or happens in a region of space-time of a finite diameter.
In the present paper we drop this requirement, not putting any
restrictions on the cardinality of $C_{\sigma\eta}$ for any $\sigma,
\eta \in \Sigma$. As a sidenote, this leads to the fact that in some
models there may be choice points which are not intuitively connected
with any splitting point. For details, see \ref{spcp} in the Appendix.

The state function assigns to each pair $\langle \mbox{a label from $\Sigma$, a point from $\reals^4$} \rangle$
an element of $P$. Colloquially, the state function tells us what happens at a certain point of the space-time
in a given scenario. \footnote{We can look at the situation from a slightly different perspective: every label
$\sigma$ is assigned a mapping $S_\sigma$ from $\reals^4$ to $P$; see also previous footnote.}

After \citet{muller_nato}, we now proceed to construct the elements of MBS version of Our World; they will be
equivalence classes of a certain relation $\leqslant_S$ on $\Sigma \times \reals^4$. For convenience, we write
the elements of $\Sigma \times \reals^4$ as $x_\sigma$ where $x \in \reals^4, \sigma \in \Sigma$. The idea is to
``glue together'' points in regions of overlap; hence the relation is defined as below:
\begin{equation}\label{S1}
x_\sigma \equiv_S y_\eta \mbox{ iff } x = y \mbox{ and } x \in R_{\sigma\eta}
\end{equation}
M\"{u}ller provides a simple proof of the fact that $\equiv_S$ is an equivalence relation on $\Sigma \times
\reals^4$; therefore we can produce a quotient structure. The result is the set $B$ being the MBS version of Our
World:
\begin{equation}\label{B1}
B := (\Sigma \times \reals^4) / \equiv_S \; \; = \{ [x_\sigma] | \sigma \in \Sigma, x \in \reals^4 \}.
\end{equation}
where $[x_\sigma]$ is the equivalence class of $x$ with respect to the relation $\equiv_S$:
\begin{equation}\label{B2}
[x_\sigma] = \{ x_\eta | x_\sigma \equiv_S x_\eta \}.
\end{equation}
Next, we define a relation $\leqslant_S$ on $B$:
\begin{equation}\label{SS1}
[x_\sigma] \leqslant_S [y_\eta] \mbox{ iff } x \leqslant_M y \mbox{ and } x_\sigma \equiv_S x_\eta
\end{equation}
which (as M\"{u}ller shows) is a partial ordering on $B$.

The goal would now be to prove that $\langle B, \leqslant_S \rangle$ is a model of BST. To do so, and in
particular to prove the prior choice principle and requirement no. 4 from definition \ref{BSTmodel}, we need to
know more about the shape of the histories in MBS - that they are the intended ones.

\subsection{The shape of MBS histories}\label{shape} We would like histories, that is:
maximal upward-directed sets, to be sets of equivalence classes $[x_\sigma]$ (with respect to $\equiv_S$) for $x
\in \reals^4$ for some $\sigma \in \Sigma$. In other words, we wish to unambiguously refer to any history by a
label from $\Sigma$, requiring one-to-one correspondence of the sets of histories and labels. This is
M\"{u}üller's [2002] Lemma 3 and our

\begin{theorem}{\label{th5}} Every history in a given MBS is of the form $h = \{
[x_\sigma]|x \in \reals^4 \}$ for some $\sigma \in \Sigma$.
\end{theorem}
The problem is that, aside from minor brushing up required by the proof of the ``right'' direction, the proof of
the ``left'' direction supplied in \cite{muller_nato} needs to be fixed as it does not provide adequate reasons
for nonemptiness of an essential intersection $\bigcap \Sigma_h (z_i)$. More on that below. Let us divide the
above theorem into two lemmas (\ref{l321} and \ref{l322}) corresponding to the directions and prove the
``right'' direction first. Until we prove the theorem we refrain from using the term ``history'' and substitute
it with a ``maximal upward-directed set'' for clarity.

\begin{lemma}{\label{l321}} If $h = \{
[x_\sigma]|x \in \reals^4 \}$ for some $\sigma \in \Sigma$ than $h$ is a maximal upward-directed subset of $B$.
\end{lemma}

\begin{proof}
Let us consider $e_1, e_2 \in h,e_1 = [x_\sigma], e_2 = [y_\sigma]$. Since $x,y \in \reals^4$ there exists a $z
\in \reals^4$ such that $x \leqslant_M z$ and $y \leqslant_M z$. Therefore $[x_\sigma] \leqslant_S [z_\sigma]$
and $[y_\sigma] \leqslant_S [z_\sigma]$, and so $h$ is upward-directed.

For maximality, consider a $g \subseteq B, g \varsupsetneq h$ and assume $g$ is upward-directed. It follows that
there exists a point $[x_\eta] \in g-h$ such that $[x_\eta] \neq [x_\sigma] \in h$. Since both points belong to
$g$ which is upward-directed, there exists $[z_\alpha] \in g$ (note that we are not allowed to choose $\sigma$
as the index at that point) such that $[x_\eta] \leqslant_S [z_\alpha]$ and $[x_\sigma] \leqslant_S [z_\alpha]$.
Therefore $x_\eta \equiv_S x_\alpha \equiv_S x_\sigma$, and so we arrive at a contradiction by concluding that
$[x_\eta] = [x_\sigma]$.
\end{proof}

The proof of the other direction is more complex and, what might be surprising, involves a topological
postulate. First, we will need a simple definition:

\begin{definition}{\label{Sigma}}
  For a given maximal upward-directed set $h$ and a point $x \in
  \reals^4$, $\Sigma_h(x) := \{ \sigma \in \Sigma | [x_\sigma] \in h
  \}$.
\end{definition}

Consider now a given maximal upward-directed set $h \subseteq B$. With every lower bounded chain $L \subset
\reals^4$ we would like to associate a topology (called ``chain topology'') on the set of $\Sigma_h(\inf(L))$.
We define the topology by describing the whole family of closed sets, which is equal to $\{ \emptyset,
\Sigma_h(\inf(L)) \} \cup \{ \Sigma_h(l) | l \in L \} \cup \{ \cap \{ \Sigma_h(l) | l \in L \} \}$. (Because $L$
is a chain it is evident that the family is closed with respect to intersection and finite union). The postulate
runs as follows:

\begin{postulate}{\label{p1}}
For every maximal upward-directed set $h \subseteq B$ and for every lower bounded chain $L \subset \reals^4$ the
``chain topology'' described above is compact.
\end{postulate}

It is easily verifiable that in such a topology $\{ \Sigma_h(l) | l \in L \}$ is a centred family of closed sets
(every finite subset of it has a nonempty intersection). Together with the above postulate we get this result:

\begin{corollary}{\label{c1}}
For every maximal upward-directed set $h \subseteq B$ and for every chain $L \subset \reals^4$, $\empty \bigcap
\{ \Sigma_h(l) | l \in L \} \neq \emptyset$.
\end{corollary}

\begin{lemma}{\label{l322}}
If $h$ is a maximal upward-directed subset of $B$ then $h = \{ [x_\sigma]|x \in \reals^4 \}$ for some $\sigma
\in \Sigma$.
\end{lemma}

The structure of the proof mimics the proof of M\"{u}ller's (see \cite{muller_nato}). It is divided into three
parts, the first and the last being reproduced here. On the other hand, the second part contains an error (as
stated above, the statement that $\bigcap \Sigma_h (z_i) \neq \emptyset$ is not properly justified) and bears on
an assumption that for every history $h$ and point $x \in \reals^4$ the set $\Sigma_h(x)$ is at most countably
infinite. We wish both to drop this assumption and correct the proof using the above topological postulate.

\begin{proof}
  Suppose that $h$ is a maximal upward-directed subset of $B$. In
  order to prove the lemma, we will prove the following three steps:

1. If for some $\sigma, \eta \in \Sigma$ both $[x_\sigma] \in h$ and $[x_\eta] \in h$, then $x_\sigma \equiv_S
x_\eta$.

2. There is a $\sigma \in \Sigma$ such that for every $\eta$, if $[x_\eta] \in h$, then $x_\eta \equiv_S
x_\sigma$.

3. With the $\sigma$ from step 2, $h = \{ [x_\sigma] | x \in \reals^4 \}$.

\medskip

\textbf{Ad. 1.} Since $h$ is maximal by assumption, there exists a $[y_\gamma] \in h$ such that $[x_\sigma]
\leqslant_S [y_\gamma]$ and $[x_\eta] \leqslant_S [y_\gamma]$. These last two facts imply that $x_\sigma
\equiv_S x_\gamma \equiv_S x_\eta$, so by transitivity of $\equiv_S$ we get $x_\sigma \equiv_S x_\eta$.

\textbf{Ad. 2.} Assume the contrary: $\forall \sigma \in \Sigma \; \exists [x_\eta] \in h, x_\eta \not\equiv_S
x_\sigma$.

Take a point $[y_\kappa] \in h$. Accordingly, $\Sigma_h(y) \neq \emptyset$.

For each scenario $\sigma_\alpha \in \Sigma_h(y)$ we define a set $\Theta_\alpha = \{ x \in \reals^4 \; | \;
\exists \eta \in \Sigma_h(y): \; [x_\eta] \in h \; \wedge \; x_{\sigma_\alpha} \not\equiv_S x_\eta \}$, which by
our assumption is never empty. Colloquially, it is a set of the points that make the scenario a wrong candidate
for the proper scenario from our lemma - the scenario ``doesn't fit'' the history at those points. For each
scenario $\sigma_\alpha$ we would like to choose a single element of $\Theta_\alpha$, and to that end we employ
a choice function $T$ defined on the set of subsets of $\reals^4$ such that $T(\Theta_\alpha) \in
\Theta_\alpha$, naming the element chosen by it as follows: $T(\Theta_\alpha) := x_\alpha$.\footnote{Bear in
mind that since $\alpha$ is a number serving just as an index for scenarios, $x_\alpha$ (like $x_\beta$ in the
line below inequality \ref{contr1}) is a point from $\reals^4$ and does not denote a point - scenario pair.}

Observe that we will arrive at a contradiction if we prove that
\begin{equation}{\label{contr1}}
\bigcap_{\sigma_\alpha \in \Sigma_h(y)} \Sigma_h(x_\alpha) \neq \emptyset
\end{equation}
(since for any $\sigma_\beta \in \Sigma_h(y) \; \sigma_\beta \notin
\Sigma_h(x_\beta)$). In order to apply our topological postulate, we
will construct a chain $L = \{ z_0, z_1,$ ... ,$ z_\omega, ... \}$ of
points in $\reals^4$. It will be lower bounded by its initial element
$z_0$. Moreover, we want it to be vertical, since this way it will (if
it does not have an upper bound itself) contain an upper bound of any
point in $\reals^4$, which will be needed in our proof.

We first define a function ``$up$'' which given two points $a, b \in \reals^4$ will produce a point $c \in
\reals^4$ such that $c$ has the same spatial coordinates as $a$ but is above $b$. In other words, if $a =
\langle a^0, a^1, a^2, a^3 \rangle \in \reals^4$, $b = \langle b^0, b^1, b^2, b^3 \rangle \in \reals^4$,
$up(a,b) := \langle a^0+(\sum\limits_1^3 (a^i - b^i)^2)^{1/2}, \; a^1, a^2, a^3 \rangle \in \reals^4$. Notice
that $up$ is not commutative.

We proceed to define the above mentioned chain $L$ in the following way:

\emph{1}.$z_0 = up(y,x_0)$.

$z_1 = up(z_0,x_1)$.

Generally, $z_{k+1} \; = \; up(z_k, x_{k+1})$.

\emph{2}. Suppose $\rho$ is a limit number. Define $A_\rho := \{
z_\beta \; | \; \beta < \rho \}$\footnote{Again: $\beta$ is just an
  index, not a scenario, so $A_\rho$ is a subset of $\reals^4$.}. As
you can see, $A_\rho$ is the part of our chain we have managed to
construct so far. We need to distinguish two cases:

a) $A_\rho$ is upper bounded with respect to $\leqslant_M$. Then it
has to have ``vertical'' upper bounds $t_0, t_1 ...$ with spatial
coordinates $t^i_n = z_0^i$ $(i = 1,2,3)$. In this case, we employ the
above defined function $T$ to choose one of the upper bounds of
$A_\rho$:
\begin{equation}
t_\rho := T( \{ t \in \reals^4 \; | \; \forall \beta < \rho \; z_\beta \leqslant_M t \wedge t^i = z_0^i (i =
1,2,3)\}).
\end{equation}
Then we put $z_\rho := up(t_\rho, x_\rho)$, arriving at the next element of our chain $L$.

b) if $A_\rho$ is not upper bounded with respect to $\leqslant_M$, then no matter which point in $\reals^4$ we
choose, it is possible to find a point from $A_\rho$ above it (since $A_\rho$ is vertical). Therefore the set
\begin{equation}
B_\rho = \{ t \in A_\rho \; | \; x_\rho \leqslant_M t \}
\end{equation}
is not empty. We put $[z_\rho] := T(B_\rho)$, arriving at the next element of our chain $L$.

Notice that in our chain it might happen that while $\alpha < \beta$, $z_\beta  \leqslant_M z_\alpha$, but $z_0$
is a lower bound of $L$. Therefore our postulate \ref{p1} applies. By employing it and corollary \ref{c1} we
infer that
\begin{equation}\label{zety}
\bigcap_{\sigma_\alpha \in \Sigma_h(y)} \{ \Sigma_h(z_\alpha) | z_\alpha \in L \} \neq \emptyset
\end{equation}
By our construction of the chain $L$, for all $\alpha$ it is true that $x_\alpha \leqslant_M z_\alpha$.
Therefore $\Sigma_h(z_\alpha) \subseteq \Sigma_h(x_\alpha)$. Thus, from \ref{zety} we immediately get
\begin{equation}
\bigcap_{\sigma_\alpha \in \Sigma_h(y)} \Sigma_h(x_\alpha) \neq \emptyset,
\end{equation}
which is the equation \ref{contr1} that we tried to show. Therefore we arrive at a contradiction and part 2 of
the proof is complete.

\textbf{Ad. 3.} We have shown that there is a scenario $\sigma \in \Sigma$ such that all members of $h$ can be
identified as $[x_\sigma]$ for some $x \in \reals^4$. What remains is to show that the history cannot
``exclude'' some regions of $\{ \sigma \} \times \reals^4$, that is: to prove that for all $x \in \reals^4,
[x_\sigma] \in h$. But in lemma \ref{l321} we have shown that $\{ [x_\sigma] | x \in \reals^4 \}$ is a maximal
upward-directed subset of $B$, so any proper subset of it cannot be maximal upward-directed.
\end{proof}

By showing lemmas \ref{l321} and \ref{l322} we have proved theorem \ref{th5}.

\subsection{The importance of the topological postulate}
So far it might seem that our topological postulate \ref{p1} is just a handy trick for proving the lemma
\ref{l322}. To show its importance we will now prove that its falsity leads to the falsity of the lemma, and
then present an example of a structure in which the lemma does not hold.

\begin{theorem}
If the postulate \ref{p1} is false, then lemma \ref{l322} is also false.
\end{theorem}

\begin{proof}
Assume that our topological postulate does not hold. Therefore there exists a maximal upward-directed set $h
\subseteq B$ and a lower bounded chain $L \subset \reals^4$ such that the chain topology is not compact. This is
by rules of topology equivalent to the fact that there is a centred family of closed sets with an empty
intersection. But all closed sets in the chain topology form a chain with respect to inclusion. Of course, if a
part of a chain has an empty intersection, a superset of the part also has an empty intersection. We infer that
\begin{equation}
\bigcap_{x \in L} \Sigma_h(x) = \emptyset
\end{equation}
from which, by definition \ref{Sigma}, we get that
\begin{equation}
\neg \exists \sigma \in \Sigma : \forall x \in L \; [x_\sigma] \in h
\end{equation}
so there is no scenario $\sigma$ such that $h = \{ [x_\sigma] | x \in \reals^4 \}$. Thus, lemma \ref{l322} is
false.
\end{proof}

\medskip
In the Appendix (section \ref{imptop}) we show a situation in which lemma \ref{l322} does not hold. The
construction resembles the $M_1$ structure from \cite{muller06:_funny}.

\subsection{BST models and MBS}
Having proven theorem \ref{th5} we can adopt M\"{u}ller's proof (from \cite{muller_nato}) of the fact that
$\langle B, \leqslant_S \rangle$ meets all the requirements in definition \ref{BSTmodel} and conclude that it is
a model of BST. We keep in mind, though, that we have introduced a new postulate \ref{p1} into the proof and
shown that it is not trivial (not always true). We will demand from the structures we would like to call
``Minkowskian Branching Structures'' to meet our topological postulate. This way, a MBS is a special kind of a
BST model: its Our World and ordering $\leqslant$ are constructed as respectively $B$ and $\leqslant_S$ as
proposed by M\"{u}ller, and furthermore our postulate \ref{p1} is true in the model.

Due to the following self-evident Fact, we have fulfilled our promise from the introduction and produced BST
models in which histories are isomorphic to Minkowski space-times.

\begin{fact}
Let $\mathcal{W} = \langle W, \leqslant_S \rangle$ be an MBS and let $h$ be a history in $\emph{W}$ of the form
$\{ [x_\sigma] \mid x \in \reals^4 \}$ for a certain $\sigma \in \Sigma$. Then
\begin{equation*}
\langle h , \leqslant_S \mid_h \rangle \cong \langle \reals^4 , \leqslant_M \rangle
\end{equation*}
by means of the isomorphism $i: h \rightarrow \reals^4$ such that $i([x_\sigma]) = x$.
\end{fact}

\section{Funny business}{\label{FBdef}}\label{sec3}
The rest of the paper concerns the funny business phenomenon in its
finitary and infinitary variants. Funny business in BST is to resemble
EPR correlations. The underlying idea is that there are two space-like
related events whose outcomes are correlated in the sense that a
combinatorially possible history is missing.  As an example, consider
a BST model of the EPR-Bohm experiment \citep{bohm51}. There are two
space-like separated measurement events $e_1$ and $e_2$, idealized to
be point-like. Each has two outcomes, `spin up' and `spin down', to be
written as, resp. $H_1\!  +$, $H_1\! -$ and $H_2\!  +$, $H_2\! -$.
Since histories with results `spin up' and `spin down' are possible,
we assume that the intersections $H_1\!  + \cap H_2\! -$ and $H_1\! -
\cap H_2\! +$ are nonempty. Yet, we put: $H_1\!  + \cap H_2\! + =
H_1\! - \cap H_2\!  - = \emptyset$, since no history with same spin
projections is possible.

Taking a clue from this example, funny business seems to require two
SLR point events $e_1, e_2$ such that for some outcomes $H_1$ of $e_1$
and $H_2$ of $e_2$: $H_1 \cap H_2 = \emptyset$. Since $e_1$ SLR $e_2$,
the two share a history. Now, this basic idea could be generalized in
two directions, giving rise to the notions of finitary funny business
and infinitary funny business. As for the former, following
\citet{belnap_nato} we allow for extended (i.e., not point-like)
events, say $A$ and $B$, require that they are SLR in the sense that
$\forall x \in A \forall y \in B \; (x \; SLR \; y)$, and we do not
postulate that the sum of the two events to be a subset of a
history.\footnote{The reason for this selection is, in the last
  instance, the existence of `nice' theorems following from this
  concept. For more, cf. footnote ? of \citet{belnap_nato}.} The
result is Belnap's notion of \emph{generalized-primary
  space-like-related modal-correlation-funny business}, i.e., g-p
s-l-r m-c funny business for short.

To obtain the infinitary version of funny business, consider a set $S$
of (not necessarily SLR) infinitely many point events and require that
any two elements of $S$ be consistent in the sense that every outcome
of the first event intersects non-emptily with every outcome of the
other; that feature obtains for any finite subset of $S$.  The funny
business consists in there being a set of outcomes of events of $S$,
one outcome for each element of $S$, the intersection of which is
empty. The resulting notion, which is closely related to {\em
  combinatorial funny business} of \citet{muller06:_funny} gives rise
to a more familiar concept if one further requires that $S$ is
pairwise SLR and is a subset of some history $h$.

To properly define funny business, we will need a few formal notions.
\begin{definition}
  $Hist$ is the set of all histories in the model.  $H_{(e)}$ is the
  set of all histories to which point event $e$ belongs.\\  For $e_1,
  e_2 \in W$, $e_1$ SLR $e_2$ iff $\exists h \in Hist:\; e_1, e_2 \in
  h$ and $e_1 \not\leqslant e_2$ and $e_2 \not\leqslant e_1$.\\  For
  $E_1, E_2 \subseteq W$, $E_1$ SLR $E_2$ iff $\forall e_1 \in
  E_1\;\forall e_2 \in E_2:\; e_1 \text{ SLR } e_2$.
\end{definition}
Next, replacing the informal notion of an event's outcome, we have a
concept of ``elementary possibility at $e$'', defined as an element of
a certain partition of $H_{(e)}$. The partition is an equivalence
relation $\equiv_e$ on $H_{(e)}$ which is to convey the sense of
``being undivided in $e$'' - sharing a point above $e$.
\begin{definition}
  Consider $h_1, h_2 \in H_{(e)}$. $h_1 \equiv_e h_2$ iff $\exists e^*
  > e$ such that $e^* \in h_1 \cap h_2$. $h_1 \perp_e h_2$ iff $h_1,
  h_2 \in H_{(e)}$ and it is not the case that $h_1 \equiv_e h_2$.
\end{definition}
The relation $\equiv_e$ is an equivalence relation on $H_{(e)}$ due to
BST postulates, as shown in \citet{belnap92}, Facts 45--46.
\begin{definition}
  If $h \in H_{(e)}$, we say that $\Pi_e \langle h \rangle \subseteq
  H_{(e)}$ is an elementary possibility (open) in $e$ iff it is the
  equivalence class of the history $h$ w.r.t. the relation $\equiv_e$.
  If $x \in W$ and $e < x$, by $\Pi_e \langle x \rangle$ we mean the
  elementary possibility in $e$ to which  history $h \in H_{(x)}$
  belongs.
\end{definition}

\noindent Following the existing literature we define $\Pi_e$ as the
set of all elementary possibilities at $e$.

Next, for a given set $S$ we will consider functions $f$ which, given
a point $e \in S$ as an argument, produce an elementary possibility
from $\Pi_e$. Colloquially speaking, function $f$ resembles a pointer,
indicating for every $e \in S$ which elementary possibility at $e$ is
selected.  Formally, a pointer function is an element of the set
$\prod_{e \in S} \Pi_e$ of product functions, defined as follows: :
\begin{equation}\label{Produkt}
  \prod_{e \in S} \Pi_e = \{ f : S \rightarrow \bigcup_{e \in S} \Pi_e : \forall e' \in S \; f(e') \in \Pi_{e'}
  \}
\end{equation}
The definitions of no (in)finitary funny business and (in)finitary
funny business run as follows:

\begin{definition}{\label{FB}}
  Assume $S \subseteq W$ and a function $ f\in \prod\limits_{e \in S}
  \Pi_e$.

  \noindent \textbf{$\langle S, f \rangle$ is not a case of finitary
    funny business} iff for any $ A_1, A_2 \subseteq S$: if $A_1$ SLR
  $A_2$ and $\bigcap \{ f(e) : e \in A_i \} \neq \emptyset$ for $i =1,
  2$, then $\bigcap \{ f(e) : e \in A_1 \cup A_2 \} \neq \emptyset$.

\noindent   \textbf{$\langle S, f \rangle$ is not a case of infinitary funny
  business} iff (1) $card(S) < \omega$ or (2) $\exists A\: (A
\subseteq_{fin} S \wedge \forall h \in Hist; \; A \not\subseteq h)$ or (3) if $\forall e, e' \in S\; (e < e'
\rightarrow f(e') \subseteq f(e))$, then $\bigcap \{ f(e) : e \in S \} \neq \emptyset$.

\noindent  \textbf{$S$ does not give rise to (in)finitary funny
    business } iff $\forall f \in \prod\limits_{e \in S} \Pi_e$
  $\langle S, f \rangle$ is not a case of (in)finitary funny business.
\end{definition}

\noindent On this definition, $\langle S, f \rangle$ is a case of finitary funny business iff there are $ A_1,
A_2 \subseteq S$ such that $A_1$ SLR $A_2$ and $\bigcap \{ f(e) : e \in A_i \} \neq \emptyset$ for $i =1, 2$ but
$\bigcap \{ f(e) : e \in A_1 \cup A_2 \} = \emptyset$. I.e., this is a case of Belnap's g-p s-l-r m-c funny
business (for details, see Appendix).

\noindent And $\langle S, f \rangle$ is a case of infinitary funny
business iff (1) $card(S) \geqslant \omega$ and (2) $\forall A\; (A
\subseteq_{fin} S \rightarrow \exists h \in Hist: \; A \subseteq h)$,
and (3) $\forall e, e' \in S\; (e < e' \rightarrow f(e') \subseteq
f(e))$ and (4) $\bigcap \{ f(e) : e \in S \} = \emptyset$. To see the
rationale underlying clause (3), note that due to that clause, there
is no infinitary funny business if for some $e, e' \in S\; (e < e'
\wedge f(e') \not\subseteq f(e))$, which entails, by properties of
elementary possibilities, that (*) $f(e') \cap f(e) = \emptyset$.  In
other words, the reason why the infinite intersection (4) is empty is
that the intersection (*) is empty.  In a similar vein, by clause (2),
we do not call it infinitary funny business if some finite subset of
$S$ is inconsistent: in this case there is nothing funny in $\bigcap
\{ f(e) : e \in S \} = \emptyset$, as this follows from the above.

The relation to \citet{muller06:_funny}'s notion of combinatorial funny business is this:

\noindent If $\langle S, f\rangle$ is a case of infinitary funny business, then $\langle S, f\rangle$
constitutes a case of combinatorial funny business. In the other direction, if $\langle S, f\rangle$ constitutes
a case of combinatorial funny business but $\langle S, f\rangle$ is not a case of finitary funny business, then
$\langle S, f\rangle$ is a case of infinitary funny business.  For details, see Appendix.

One might find the above definitions not completely intuitive or even objectionable, and to some extend we share
this feeling. For instance, in what follows, while discussing finitary and infinitary funny business, we are
concerned with only such $S$ that are pairwise SLR and a subset of a history. So, the definitions given above
are too general for our purposes.  Despite these disadvantages, we assume them since they are closely related to
the extant definitions, and there are some odd structures arising from $S$ that is neither pairwise SLR nor a
subset of a history. For brevity, from now on instead of ``finitary funny business'' we will usually write
``FINFB'' and instead of ``infinitary funny business'' we will usually write ``INFFB''. We will also say that NO
FINFB (NO INFFB) is true in a BST model ${\mathcal W} = \langle W, \leqslant \rangle$ meaning that no $S
\subseteq W$ gives rise to FINFB (INFFB).

\subsection{$M_2$}\label{M2}
\cite{muller06:_funny} introduced a certain BST structure named $M_2$, in which FINFB was absent, whereas INFFB
was present. We will now briefly reproduce their definition, because it is an interesting example of funny
business and we will use it in our theorems. For a detailed discussion and a proof that $M_2$ is a BST model
with the above properties, see the mentioned paper.

$M_2$ is a pair $\langle W, \leqslant \rangle$. $W$ is a union of four sets: $W_0 = (-\infty,0]$, $W_1 = (0,1]
\times \mathbb{N}$, $W_2 = (1,2) \times \mathbb{N} \times \{0,1\}$ and $W_3 = [2,\infty) \times \mathbb{F}$
where $\mathbb{F}$ is the set of all functions $g:\mathbb{N} \rightarrow \{0,1\}$ such that for only finitely
many $n \in \mathbb{N}$, $g(n) = 0$.

The strict partial ordering $<$ is the transitive closure of the following four relations:
\begin{itemize}
\item For $e,e_1$ from the same $W_i$: $e < e_1$ iff the first
  coordinate of $e$ is smaller than that of $e_1$ and the other
  coordinates are the same.
\item $x < (y,n)$ for every $x \in W_0$ and $(y,n) \in W_1$.
\item For $(x,n) \in W_1$ and $(y,m,i) \in W_2: (x,n) < (y,m,i)$ iff
$n=m$.
\item For $(x,n,i) \in W_2$ and $(y,g) \in W_3: (x,n,i) < (y,g)$ iff
  $(g(n)=i$.
\end{itemize}
The non-strict companion $\leqslant$ of $<$ is defined as usual: $e \leqslant e'$ iff $e < e'$ or $e =e'$. The
structure $M_2$ has a countable set $Hist$ of histories and also a countable set of binary choice points $S = \{
\langle 1, n\rangle : n \in \mathbb{N} \}$. Moreover, there is one-to-one correspondence between $Hist$ and
$\mathbb{F}$, which allows us to identify values of product functions (i.e., elementary possibilities) with
certain subsets of $\mathbb{F}$. \label{hisM2} At each point $e = \langle 1, n \rangle$ there are two elementary
possibilities, both of the form $\{g \in \mathbb{F} \mid g(n) = b\}$, where $b$ is $0$ or $1$. Taking now a
product function $f \in \prod\limits_{e \in S} \Pi_e$ such that $f(\langle 1 , n\rangle) = \{g \in \mathbb{F}
\mid g(n) = 0\}$, it is easy to see that $\langle S, f\rangle$ is a case of INFFB since in $\mathbb{F}$ there is
no function $z$ such that $z(x)=0$ for every $x \in {\mathbb N}$. On the other hand, there is no case of FINFB
in $M_2$ (see {\citet{muller06:_funny}). Thus, $M_2$ is just a case in
  point: it has INFFB that does not involve any case of FINFB. In a
  due time, we will ask if the structure can be `converted' into a
  MBS. At this stage, let us note some `strange' features of $M_2$.
  First, in $M_2$ a point above some two choice points is always above
  an infinite number of choice points. Also, in $M_2$ one can define a
  certain `odd' subset $X$, of which our Postulate~B (to be introduced
  later) is true. Namely,
\begin{equation}
X := \{ \langle \frac{3}{2}, n , 0\rangle \mid n \in {\mathcal N}\} \label{setX}
\end{equation}
$X$ is strange because it is not a subset of any history, yet, every finite subset of $X$ is contained in some
history.

\subsection{Results}\label{results}

One might expect that there are cases of INFFB that involve FINFB: indeed the theorem below justifies this
intuition and gives it a precise reading.

\begin{theorem}
  If $\langle S, f\rangle$ is a case of FINFB and for some history
  $h_S$: $S \subseteq h_S$, then there are $S' \subseteq W$ and $f' \in
  \prod\limits_{e \in S'} \Pi_e$ such that $\langle S', f'\rangle$ is
  a case of INFFB.
\label{fin2inf}
\end{theorem}

\noindent Proof: By the assumption, there are $ A, B \subseteq S$ such that (1) $A \; SLR\; B$ and (2) $h_A \in
\bigcap \{ f(e) : e \in A \}$, (3) $h_B \in \bigcap \{ f(e) : e \in B \}$ but (4) $\bigcap \{ f(e) : e \in A
\cup B \} = \emptyset$.  Clearly, (5) $A \subseteq h_A \cap h_S$ and (6) $B \subseteq h_B \cap h_S$. If $A\cup
B$ is infinite, put: $S' = A \cup B$ and $f' = f_{\mid A\cup B}$, and go to ($\dagger$). If $A\cup B$ is is
finite, define now $S' := \{x \in h_S\mid \neg \exists e_1 \in A, e_2 \in B\; (x> e_1 \wedge x > e_2)\}$.

Let us first argue that $S'$ is infinite. Pick $a$: a maximal element of $A$ and $b$: a maximal element of $B$;
since both share $h_S$, there is $y$ such that $y > a \wedge y>b$. Consider next a chain $l_a$ from $y$ to $a$,
such that $inf(l_a) = a$. If (7) $\forall x\; (x > a \rightarrow x >b)$, then $l_a$ is lower bounded by $b$.
Yet, since $a$ is the infimum, i.e., the greatest lower bound of $l_a$, it must be that $a \geqslant b$, which
contradicts (1). Hence, since (7) cannot be true, there is an $x$ such that $x< y$ and $x > a$ but $x \not> b$.
BST's postulate of density implies that $Z := \{z \in l_a \mid z < x\}$ is infinite; since $Z \subseteq S'$, $S'
$ is infinite as well.

As for the next condition of INFFB, since $S' \subseteq h_S$, it must be that $\forall A \subseteq_{fin} S'\; A
\subseteq h_S$.

We define $f' \in \prod\limits_{e \in S'} \Pi_e$:

\begin{equation*}
    f'(e) = \begin{cases}
    f(e) \text{\; if \;} e \in A \cup B\\
    \Pi_e\langle h_A\rangle  \text{\; if \;}  e \not\in A \cup B \text{ and }\exists x\in A (e < x)\\
    \Pi_e\langle h_B \rangle \text{\; if \;} e \not\in A \cup B \text{
      and } \exists x\in B (e < x)\\
    \Pi_e\langle h_S \rangle \text{\; otherwise \;} \\ \end{cases}
\end{equation*}

\noindent
\noindent ($\dagger$) By the definitions of $S'$ and $f'$ and (1)-(6) we have: $\forall e, e' \in S'\; (e < e'
\rightarrow f'(e') \subseteq f'(e))$. By (4), since $A \cup B \subseteq S'$, we get $\bigcap \{ f'(e) : e \in S'
\} = \emptyset$. \quad $\square$.

\medskip
\noindent Our title question, however, is just the opposite: are there cases of INFFB that do not involve FINFB?
Before we turn to our main theorems, let us first show some simpler facts:

\begin{corollary}
  Suppose that $A\subseteq S$ is finite and pairwise $SLR$. Then, if
  $S$ does not give rise to FINFB, $\bigcap\limits_{e \in A} \{ f(e)
  \} \neq \emptyset$ for any $f \in \prod\limits_{e \in S} \Pi_e$.
\label{useful}
\end{corollary}
The corollary stems from the fact that any finite set is a union of a finite family of singletons.

\begin{theorem}
  Assume that $S$ is an infinite set of pairwise SLR points such that
  for some history $h$, $S \subseteq h$.

If there exist sets $A_1, A_2$ such that $A_1 \cup A_2 = S$ and none of them gives rise to INFFB, then (if $S$
gives rise to INFFB, then S gives rise to FINFB).
\end{theorem}
\begin{proof}
  From the first antecedent we get that $\forall f \in \prod\limits_{e
    \in A_1} : \bigcap \{ f(e): e \in A_1 \} \neq \emptyset$ and a
  similar result for $A_2$. From the second antecedent we get that
  $\exists g \in \prod\limits_{e \in S}: \bigcap \{ f(e): e \in S \} =
  \emptyset$. We can of course think of the function $g$ defined on
  $S$ as a union of two functions defined respectively on $A_1$ and
  $A_2$. Thus, we see that $\langle S, g \rangle$ constitute a case of
  FINFB because $\bigcap \{ g(e) : e \in A_i \} \neq \emptyset$ and
  $\bigcap \{ g(e) : e \in A_2 \} \neq \emptyset$ while $\bigcap \{
  g(e) : e \in A_1 \cup A_2 \} = \emptyset$. Therefore $S$ gives rise
  to FINFB.
\end{proof}

\bigskip
The above theorem yields us the following simple corollary:

\begin{corollary}
  Assume $S$ is an infinite set of pairwise SLR points such that for
  some history $h$, $S \subseteq h$. Then, if $S$ does not give rise
  to FINFB and there exists a cofinite subset of $S$ which does not
  give rise to INFFB, then the whole set S does not give rise to
  INFFB.
\end{corollary}

\medskip
Turning to our main theorems  characterizing INFFB, we will now introduce two postulates and prove a few
theorems about and show  how they relate to FINFB and INFFB.
\begin{postulate}\label{posA}
(\textit{\textbf{Postulate A}}) There exist 1) a set $S \subseteq W$ which is an infinite set of pairwise SLR
points such that for some history $h$ $S \subseteq h$ and 2) a function $ f\in \prod\limits_{e \in S} \Pi_e$
such that
\begin{eqnarray*}\label{POS1}
\lefteqn{\exists e \in S \; \forall h \in Hist \; \forall x \in W:} \\
& & ( x \notin h \; \vee \; \neg (x > e) \; \vee \; h \notin f(e) \; \vee \; \exists e_1 \in S (h \notin f (e_1)
\wedge \neg (x \; SLR \; e_1)))
\end{eqnarray*}
\end{postulate}

The motivation for this postulate comes from a certain structure,
called $M_1$, that \citet{muller06:_funny} introduced. In this
structure, one tries `by hand' to prohibit the existence of a certain
history, by this means producing a case of INFFB, without there being
a case of FINFB.  This procedure, however, fails if Postulate~A is
false. In this case, a seemingly excluded history gets re-inserted
``by force'' by Kuratowski-Zorn Lemma.  Namely, the falsity of
Postulate~A ensures the existence of a certain function that can be
used to produce a chain of events, which extends, by Kuratowski-Zorn
lemma to the seemingly excluded history. More precisely, if Postulate
A is false, then for any infinite pairwise SLR set $S$ such that for
some history $h$ $S \subseteq h$ and for any function $ f\in
\prod\limits_{e \in S} \Pi_e$ we can define a function $F : S
\rightarrow Hist \times W$ in the following way ($e \in S$):
\begin{equation}\label{funkcjaF}
F(e) := \langle h , x \rangle : (x>e \wedge x \in h \wedge h \in f(e) \wedge \forall_{e^\prime \in S} (h \notin
f(e^\prime) \Rightarrow e^\prime \; SLR \; x))
\end{equation}
(Of course many different functions meeting this requirement might
exist as there might be many equally good candidates for $h$ and $x$
such that for a given $e$ $F(e) = \langle h, x \rangle$. What is
important for us is that, when Postulate A is false, such functions
\textit{do} exist; we will just choose one.) Thus, we assume
Postulate~A in order Kuratowski-Zorn lemma {\em not} to produce
unwanted histories.

As for the second postulate, it relates to structure $M_2$ and a question why it contains a case of INFFB. Our
tentative diagnosis is that in $M_2$ one can define a certain `odd' subset $X$ (see Equation~\ref{setX}), of
which the postulate below is true:

\begin{postulate}\label{posB}
(\textit{\textbf{Postulate B}})
\medskip
There is a set $X \subseteq W$ such that

\begin{enumerate}
\item[(a)] for any $A \subseteq_{fin} X$, there is a history $h$: $A \subseteq h$;
\item[(b)] there is no history $h$ such that $X \subseteq h$.
\end{enumerate}
\end{postulate}

The theorems we will show are summarized in the list below:

\begin{enumerate}
\item (Theorem \ref{tt1})\; $Postulate A \Rightarrow $INFFB
\item (Theorem \ref{tt2}) \; $Postulate B \wedge $ NO FINFB $ \Rightarrow$ INFFB
\item (Theorem \ref{tt3}) \; Given that the BST model has space-time
  points,

NO FINFB $\wedge \neg(Post. A) \wedge \neg(Post. B) \Rightarrow$ NO INFFB
\item (Theorem \ref{tt4}) \; $Postulate A \nRightarrow$ FINFB
\item (Theorem \ref{tt5}) \; $\neg(Postulate A) \wedge Postulate B
   \nRightarrow$ FINFB
\end{enumerate}

As for ``space-time points'' mentioned in theorem \ref{tt3}, in its
proof we want to be able to say that something happens ``in the same
space-time point'' in different histories. A triple $\langle W,
\leqslant, s \rangle$ is a ``BST model with space-time points''
(BST+S) iff $\langle W, \leqslant \rangle$ is a BST model and $s$
(from the expression ``space-time point'') is an equivalence relation
on $W$ such that 1) for each history $h$ in $W$ and for each
equivalence class $s(x)$, $x \in W$, the intersection $h \cap s(x)$
contains exactly one element and 2) $s$ respects the ordering: for
equivalence classes $s(x), s(y)$ and histories $h_1, h_2$, $s(x) \cap
h_1 = s(y) \cap h_1$ iff $s(x) \cap h_2 = s(y) \cap h_2$, and the same
for ``$<$'' and ``$>$''. As M\"{u}ller shows in \cite{mueller05},
not every BST model can be extended to a BST+S model, so our theorem
is not as general as we would ideally prefer.

\begin{theorem}\label{tt1}
  Suppose Postulate A is true due to some $S \subseteq W$ and $f \in
  \prod\limits_{e \in S} \Pi_e$. Then $\langle S, f \rangle$ is a case
  of INFFB.
\end{theorem}
\begin{proof}
  Since by the assumption $S$ is infinite, pairwise SLR, and for some
  history $h$: $S \subseteq h$, we have: (1) $card(S) \geqslant
  \omega$ and (2) $\forall A\; (A \subseteq_{fin} S \rightarrow
  \exists h \in Hist: \; A \subseteq h)$, and (3) $\forall e, e' \in
  S\; (e < e' \rightarrow f(e') \subseteq f(e))$. We need thus to show
  (4) $\bigcap \{ f(e) | e \in S \} = \emptyset$. For {\em reductio}
  assume $\bigcap \{ f(e) | e \in S \} \neq \emptyset$.  Hence, there
  must be a history (a) $h^* \in \bigcap \{ f(e) | e \in S \}$.
  Suppose $e^* \in S$ is one of the points of which the existential
  formula of Postulate A is true.  Since it follows that $h^* \in
  f(e^*)$, it is true for $e^*$ that
\begin{equation}
\forall x \in W (x \notin h^* \; \vee \; \neg(x > e^*) \; \vee \; \exists e_1 \in S (h^* \notin f(e_1) \wedge
\neg (x \; SLR \; e_1))).
\end{equation}
Again, since $h^* \in f(e^*)$ and there are no maximal elements in the model (see BST postulate~2 of
Definition~\ref{BSTmodel}), we can find a point $x^*$ such that $x^* > e^*$ and $x^* \in h^*$. In other words,
for this $x^*$ two elements of the above alternative are false - so the third one must be true. But it also is
false, since one of the conjuncts is always false: namely, because of (a) it can't be true for any $e_1 \in S$
that $h^* \notin f(e_1)$. So the whole alternative is false for $x^*$, and thus we arrive at a contradiction.
Therefore $\bigcap \{ f(e) | e \in S \} = \emptyset$ so $\langle S, f \rangle$ constitute a case of INFFB.
\end{proof}

\bigskip We will now establish that INFFB follows from Postulate B
together with NO FINFB. Suppose that Postulate B is true due to a
certain set $X$. Our goal is to find a set $S$ and a function $f$ such
that $\langle S, f \rangle$ is a case of INFFB.

\begin{theorem}\label{tt2}
  Suppose that in a BST model ${\mathcal W} = \langle W, \leqslant
  \rangle$ Postulate~B is true and no $S \subseteq W$ gives rise to
  FINFB. Then there exists a case of INFFB in the model, i.e., there
  exists a set $S \subseteq W$ and a function $f \in \prod\limits_{e
    \in S} \Pi_e$ such that $\langle S, f \rangle$ is a case of
  INFFB.
\end{theorem}

\noindent Proof: Let $X \subseteq  W$ be a set of which Postulate~B is true. Define:
  \begin{gather}
C(x) := \{e \in W\mid \exists h (h \in Hist \wedge h \perp_e H_{(x)})\}\\
S := \bigcup_{x\in X} C(x)\label{def:S}
  \end{gather}

  \noindent The assumption of NO FINFB and a fact about location of
  choice points (see Appendix, Fact~\ref{nbtheorem}) entail (a)
  $\forall e (e \in C(x) \rightarrow e<x)$, so it makes sense to write
  $\Pi_e\langle x \rangle$ if $e \in C(x)$.  We thus tentatively
  define the product function $f$ on $S$:

\noindent $f(e) := \Pi_e\langle x \rangle$ iff $e \in C(x)$.

\noindent To check that this is indeed a good definition, we need to prove that if (b) $e \in C(x) \cap C(y)$
and $x \neq y$, then $\Pi_e\langle x \rangle = \Pi_e\langle y \rangle$. Assume to the contrary that (c)
$\Pi_e\langle x \rangle \neq \Pi_e\langle y \rangle$. Then, since by (a) and (b): $e < x \wedge e <y$, it must
be by (c) that (d) $H_{(x)} \perp_e H_{(y)}$. But, by Postulate~B~(a), there must be a history $h^*$ such that
$\{x, y \} \subseteq h$, and hence $h^* \in H_{(x)}$ and $h^* \in H_{(y)}$, which contradicts (d).

Turning to the conditions of INFFB, we will first argue  that\\ (f)
$\forall e, e' \in S\; (e < e' \rightarrow f(e') \subseteq f(e))$.\\
If $e, e' \in C(x)$, then by the definition of $f$: $f(e) = f(e')$. So, let $e \in C(x)$, $e' \in C(y)$ and $x
\neq y$. By Postulate~B~(a), $\exists h_{xy} \in H:\; \{x, y\} \subseteq h_{xy}.$. Accordingly, $h_{xy} \in
\Pi_e\langle h_{xy}\rangle = \Pi_e\langle x \rangle = f(e)$ and $h_{xy} \in \Pi_{e'}\langle h_{xy}\rangle =
\Pi_{e'}\langle y \rangle = f(e')$.  Accordingly, $\Pi_{e}\langle x \rangle \cap \Pi_{e'}\langle y \rangle \neq
\emptyset$. Moreover, since $e<e'$, $\Pi_{e'}\langle y \rangle \subseteq \Pi_{e}\langle x \rangle \neq
\emptyset$, i.e., $f(e') \subseteq f(e)$.

Next, we prove the following:

\noindent
(g) If $h \in \bigcap\{f(e)\mid e \in S\}$, then $X \subseteq h$.\\
Indirectly, let for some $h$: $h \in \bigcap\{f(e)\mid e \in S\}$ and assume that there is $x \in X$ such that
$x \not\in X$. Take any $h_x$ such that $x \in h_x$. Clearly, $x \in h_x / h$, so by PCP: $\exists e \in W:\; (h
\perp_e h_x \wedge e < x)$, from which (i) $h \perp_e H_{(x)}$ follows. Thus, $e \in C(x)$. By the assumption,
$h \in f(e) = \Pi_e \langle x \rangle$, and hence $h \equiv_e H_{(x)}$. This contradicts (i), however.

Now, Postulate~B~(b) says that $\neg \exists h\in Hist\; X \subseteq
h$, so (g) implies that\\
(j) $\bigcap\{f(e)\mid e \in S\} = \emptyset$.

 We next prove that\\ (k) if $A\subseteq_{fin} S$, then $\bigcap\{f(e)\mid e \in
 A\} \neq \emptyset$.\\ If $A$ is finite, then there is a set $A^*$ of
 maximal elements of $A$. Clearly, $A^*$ is pairwise SLR. Hence, since
 no $S \subseteq W$ gives rise to FINFB, it follows by
 Corollary~(\ref{useful}) that \\
 (l) $\bigcap\{f(e)\mid e \in A^*\} \neq \emptyset$.
 By the construction, if $e \in A/A^*$, then there is $e' \in A^*$ such
 that $e < e'$. By (f) then, $f(e') \subseteq f(e)$. This and (l)
 entail  that $\bigcap\{f(e)\mid e \in A\} \neq \emptyset$.

 Importantly, (k) has two sought-for consequences:\\ (m) if
 $A\subseteq_{fin} S$, then $\exists h \in Hist\; (A \subseteq h)$ and
 \\(n) $card(S) \geqslant \omega$.  Otherwise, by (g) and (k) there
 would be a history $h$ such that $X \subseteq h$, which contradicts
 Postulate~B~(b).

 To see that $\langle S, f \rangle$ is a case of INFFB, we need to
 show that (1) $card(S) \geqslant \omega$ and (2) $\forall A\; (A
 \subseteq_{fin} S \rightarrow \exists\! h \in Hist: \; A \subseteq
 h)$, and (3) $\forall e, e' \in S\; (e < e' \rightarrow f(e')
 \subseteq f(e))$ and (4) $\bigcap \{ f(e) : e \in S \} =
 \emptyset$.  Yet, we already established these conditions: (1) is
 (n), (2) is (m), (3) is (f) and (4) is (j). \quad $\square$

 \medskip Note that $S$, as constructed in the proof above, gives rise
 to INFFB but it needs neither to be pairwise SLR, nor a subset of a
 history.  Thus, it might be that funny business generated by
 Postulate~B is even stranger than expected, as for instance $S$ might
 have no maximal elements.  To secure a more familiar INFFB, i.e.,
 such that $S$ is pairwise SLR and a subset of a history, we need
 another requirement, called Supplement, which refers to $S$ from the
 proof above:

\medskip
 \noindent {\em Supplement} Every chain in $S$ is upper bounded and
 for some $h^* \in Hist$: $S \subseteq  h^*$.

\medskip
\noindent We leave it to the reader to show that if Postulate~B and {\em
  Supplement} are true and no $S \subseteq W$ gives rise to FINFB,
then there exists a set $S^*$ that is pairwise SLR and a subset of history and gives rise to INFFB. The set in
question can be defined as:

\[
S^* = \{sup_{h^*}(l)\mid l \text{ is a maximal chain in } S\}
\]

\noindent where $S$ is defined in (\ref{def:S}) and $h^*$ is the history to which {\em Supplement} refers. The
sought-for product function $f'$ on $S^*$ should be defined for such $e$ that $e \in S^*/ S$ as well. Thus,

\begin{equation*}
    f'(e) :=
\begin{cases} \Pi_e\langle x \rangle \text{  iff } e \in C(x);\\
\Pi_e\langle h^* \rangle \text{  iff } e \in S^*/S. \end{cases}
\end{equation*}


\bigskip It is now time to prove our main theorem. 

\begin{theorem}\label{tt3}
  Suppose ${\mathcal W} = \langle W, \leqslant \rangle$ is a BST+S
  model and no $S \subseteq W$ gives rise to FINFB. Suppose further
  that both Postulates A and B are false in ${\mathcal W}$.  Then no
  infinite set $S$ of pairwise SLR points such that for some history
  $h$: $S \subseteq h$ gives rise to INFFB.
\end{theorem}
\begin{proof}
  Suppose that $S$ is infinite, pairwise SLR and a subset of a
  history. Since $\langle S, f \rangle$ not being a case of INFFB is
  equivalent to the disjunction of four conditions, it suffices to to
  show that one of these conditions obtains. That is, we will prove
  that for any product function $f$ on $S$, (a) $\bigcap \{f(e) : e \in S
  \} \neq \emptyset$.

  Consider $S$ as naturally indexed by its cardinality.  Since
  Postulate A is false, there is a function $F : S \rightarrow Hist
  \times W$ defined as in (\ref{funkcjaF}).  Take $e_0 \in S$. For
  some $x_0 \in W$ and $h_0 \in Hist$ we have that $F(e_0) = \langle
  h_0, x_0 \rangle$.  Consider $S_0 := \{ e \in S : h_0 \in f(e)
  \wedge x_0 > e \}$. If $S_0 = S$, we have completed the proof and
  $h_0$ is the desired history, since then $\forall e\in S_0\; (h_0 \in
  f(e))$.

  Otherwise, the construction guarantees that $x_0 \; SLR \; (S /
  S_0)$. Namely, if $e \in S / S_0$ and $h_0 \not\in f(e)$, then by
  the definition of $F$ (see eq.~(\ref{funkcjaF})), $x_0$ SLR $e$.
  And, if (b) $h_0 \in f(e)$, it cannot be that $e < x_0$, because
  then $e \in S_0$. It cannot be that $e > x_0$, either, since this
  implies $e > e_0$. By (b): $e \in h_0$, and from the definition of
  $F$: $x_0 \in h_0$, and hence $x_0$ SLR $e$.

  Take then a point from $S / S_0$ (say, a point $e_i$ such that $i$
  is the minimal index in the set of indexes of points from $S / S_0$)
  and call it $e_1$. So, for some $x_1^\prime \in W$ and $h_1^\prime
  \in Hist$ we have that $F(e_1) = \langle h_1^\prime, x_1^\prime
  \rangle$. From NO FINFB (applied to the SLR set $\{x_0, e_1\}$) we
  get that $H_{(x_0)} \cap \Pi_{e_1}\langle h_1^\prime\rangle \neq
  \emptyset$ so there is a history $h_1$ belonging to the
  intersection. Clearly, $x_0 \in h_1$.  Since ${\mathcal W}$ is by
  assumption a BST+S model, we can take a point $x_1 := s(x_1^\prime)
  \cap h_1$.  Accordingly, $\{x_0, x_1\} \subseteq h_1$. We define
  $\Sigma_1 := \{ x_0, x_1 \}$ and $H_{\Sigma_1}:= \{h \in Hist \mid
  \Sigma_1 \subseteq h\}$.  Take $S_1 := \{ e \in S / S_0 : h_1 \in
  f(e) \wedge x_1 > e \}$.  On the occasion that $S = S_0 \cup S_1$ we
  have completed the proof and $h_1$ is the desired history. For, we
  have that $\forall e\; (e \in S_0 \cup S_1 \rightarrow h_1 \in
  f(e))$.  If $S \neq S_0 \cup S_1$, we continue similarly with a
  point $e_2 \in S/ (S_0 \cup S_1)$.

  The above two steps should give us an idea of what to do while
  moving from $e_k$ to $e_{k+1}$. Suppose we finished the $k$-th step
  and accordingly we have set $S_k$, history $h_k$ and set $\Sigma_k
  \subseteq  h_k$. If $S/ \bigcup\limits_{0 \leqslant i \leqslant k} S_i
  \neq \emptyset$, the theorem is not proved yet, so we take a point
  from $S /\bigcup\limits_{0\leqslant i \leqslant k} S_i$ and label it
  $e_{k+1}$. So, for some $x_{k+1}^\prime \in W$ and $h_{k+1}^\prime
  \in Hist$ we have that $F(e_{k+1}) = \langle h_{k+1}^\prime,
  x_{k+1}^\prime \rangle$. %
  From NO FINFB (applied to set $\Sigma_k \cup \{e_{k+1}\}$ as
  $\Sigma_k \text{ SLR } \{e_{k+1}\}$) we get that $H_{\Sigma_k} \cap
  \Pi_{e_{k+1}}\langle h_{k+1}^\prime \rangle \neq \emptyset$ so there
  is a history $h_{k+1}$ belonging to the intersection. Take $x_{k+1}
  := s(x_{k+1}^\prime) \cap h_{k+1}$ and put $\Sigma_{k+1} = \Sigma_k
  \cup \{x_{k+1} \}$. Of course $\Sigma_{k+1} \subseteq h_{k+1}$.
  Define $S_{k+1} := \{ e \in S/ \bigcup\limits_{0 \leqslant i
    \leqslant k} S_i : h_{k+1} \in f(e) \wedge x_{k+1} > e \}$. On the
  occasion that $S = \bigcup\limits_{0 \leqslant i \leqslant k+1} S_i$ we have
  completed the proof and $h_{k+1}$ is the desired history. If not, we
  continue similarly with a point $e_{k+2} \in S/ \bigcup\limits_{0
    \leqslant i \leqslant k+1} S_i$.

  Let us now move to the limit case. Consider the set
  $\bigcup\limits_{k < \omega} \Sigma_k$. It possesses the following
  properties:
\begin{itemize}

\item For every finite subset $A \subseteq_{fin} \bigcup\limits_{k <
    \omega} \Sigma_k$ there exists a history $h: A \subseteq h$ (since
  it is finite, $A$ has to be a subset of $\Sigma_k$ for some $k <
  \omega$, and so $A \subseteq h_k$);
\item It is infinite (since $\forall_{i,j} (i \neq j \Rightarrow
  \Sigma_i \neq \Sigma_j))$.
\end{itemize}

Therefore, the set is of the kind that Postulate B speaks about. Since we assumed its negation, we infer that
there is a history $h^* \in Hist$ such that $\bigcup\limits_{k < \omega} \Sigma_k \subseteq h^*$. If $S =
\bigcup\limits_{k < \omega} S_k$, the theorem is proved and $h^*$ is the desired history.

Suppose however that $S / \bigcup\limits_{k < \omega} S_k \neq
\emptyset$.  Take a point $e_\omega \in S-\bigcup\limits_{k < \omega}
S_k$. So, for some $x_\omega^\prime \in W, h_\omega^\prime \in Hist$
it is so that $F(e_\omega) = \langle h_\omega^\prime, x_\omega^\prime
\rangle$. Consider sets $A_1 := \{ e_i : 0 \leqslant i < \omega \}$
and $A_2 := \{ e_\omega \}$ From the construction it follows that
there are histories $h^*$ and $h'_{\omega}$ such that $h^* \in
\bigcap\limits_{e \in A_1} f(e)$ and $h_\omega^\prime \in
\bigcap\limits_{e \in A_2} f(e)$. So, by NO FINFB (applied to $A_1
\cup A_2$), $\bigcap\limits_{e \leqslant \omega} f(e) \neq \emptyset$,
so there is a history $h_\omega$ belonging to the intersection. Put
$x_\omega := s(x_\omega^\prime) \cap h_\omega$ and let $\Sigma_\omega
:= \{ x_\omega \} \cup \{ s(x) \cap h_\omega : x \in \bigcup\limits_{k
  < \omega} \Sigma_k \}$. Let $S_\omega := \{ e \in
S/\bigcup\limits_{k < \omega} \Sigma_k : h_\omega \in f(e) \wedge
x_\omega > e \}$.

If $S = \bigcup\limits_{k \leqslant \omega} S_k$ we have completed the proof and $h_\omega$ is the desired
history.  If not, we continue similarly with points from $S / \bigcup\limits_{k \leqslant \omega} S_k$. Since we
have given instructions on what to do with e point $e_i$ whether $i$ is a limit number or not (the above case
with $\omega$ is easily generalized), we are bound to arrive at a desirable history $h \in \bigcap \{f(e) : e
\in S \}$.
\end{proof}

The last two theorems in this section are to show that the first two theorems from the list above are not
useless: since FINFB leads to INFFB, we need to make sure that neither Postulate A nor Postulate~B yields FINFB.

\begin{theorem}\label{tt4}
  If a set $S \subseteq W$ is an infinite set of pairwise SLR points
  such that for some history $h$ $S \subseteq h$ and a function $f\in
  \prod\limits_{e \in S} \Pi_e$ satisfies this condition:
\begin{eqnarray*}\label{Ppos1}
\lefteqn{\exists e \in S \; \forall h \in Hist \; \forall x \in W:} \\
& & x \notin h \; \vee \; \neg (x > e) \; \vee \; h \notin F(e) \; \vee \; \exists e_1 \in S (h \notin f (e_1)
\wedge \neg (x \; SLR \; e_1))
\end{eqnarray*}
then it does not follow that $\langle S, f \rangle$ is a case of FINFB.
\end{theorem}
\textit{Proof sketch.} Our example will take place in a MBS. Unfortunately, during the construction we have run
into similar problems as with theorem \ref{LW1}: namely, we can present a proper set if we restrict ourselves to
$\reals^2$, while the $\reals^4$ case involves an intuitive extension of our idea which unfortunately would be
formally painful. Thus we will show the $\reals^2$ case. The second coordinate is spatial. (By ``$(a,b)$'' we
will sometimes mean ``a point in $\reals^2$''or ``a segment of $\reals$'', but it will always be clear from the
context.)

Let $S_1 = \{ (0,x) \in \reals^2: x \in (0,1) \}$ be a dense segment
of splitting points. Suppose all choice points generated by $S_1$ are
binary and label one possibility ``0'' and the other ``1''. Assume
that each scenario from $\Sigma$ corresponds to a history belonging to
only a finite number of ``0''-possibilities (in harmony with lemma
\ref{l322}, there are no other histories). Put $B :=(\Sigma \times
\reals^2) / \equiv_S$.

The set of choice points generated by $S_1$ will be called $S$. $S = \{ [(0,x)_\sigma] : x \in (0,1), \sigma \in
\Sigma \}$. Consider a function $f \in \prod\limits_{e \in S} \Pi_e$ such that
\[
f([(0,x)_\sigma]) =\left\{\begin{array}{llll}1 & \textrm{if $x \geqslant 1/2$}\\
0 & \textrm{if $x < 1/2$}\\\end{array}\right.
\]
The point that will make Postulate A true is $[(0,1/2)_\sigma]$. It is because it is true that
\begin{eqnarray*}\label{ppp}
\lefteqn{\forall x \in W \; \forall h \in Hist:} \\ & & (x \in h \wedge h \in f([(0,1/2)_\sigma]) \wedge x >
[(0,1/2)_\sigma]) \Rightarrow \exists e_1 \in S (h \notin f(e_1) x > e_1)
\end{eqnarray*}
which we arrive at by transforming Postulate A. And the above is true because any point above $[(0,1/2)_\sigma]$
is also above an infinite number of points $[(0,x)_\sigma]$ such that $x \in (0,1/2)$. Any history has to belong
to the ``1''-possibility in some of those points, contrary to what function $f$ dictates. Now we have to show
that $\langle S, f \rangle$ do not constitute a case of FINFB. Consider $A,B \subseteq S$. If $\{
[(0,x)_\sigma]: x \in (0,1/2)\} \cap A$ is infinite or $\{ [(0,x)_\sigma]: x \in (0,1/2)\} \cap B$ is infinite,
then from our assumption about the histories in our model we infer that $\bigcap \{ f(e) : e \in A \} =
\emptyset$ (resp. $\bigcap \{ f(e) : e \in A \} = \emptyset$), so the antecedent from the definition of NO FINFB
is false. In the other case, if both $\{ [(0,x)_\sigma]: x \in (0,1/2)\} \cap A$ and $\{ [(0,x)_\sigma]: x \in
(0,1/2)\} \cap B$ are finite, then $\{ [(0,x)_\sigma]: x \in (0,1/2)\} \cap ( A \cup B )$ is finite, therefore
(again, by our assumption about the histories in the model) $\bigcap \{ f(e) : e \in A \cup B \} \neq
\emptyset$, so the consequent from the definition of NO FINFB is false. Therefore, $\langle S, f \rangle$ do not
constitute a case of FINFB.

\begin{theorem}\label{tt5}
  Suppose Postulate A is false, whereas Postulate~B is true in our
  model and $X$ is the set whose existence is entailed by Postulate B.
  It does not follow that $S$ gives rise to FINFB.
\end{theorem}
\textit{Proof by observation}: $M_2$ provides us with an appropriate example (see section~\ref{M2}). Set $X$, as
defined in Equation~(\ref{setX}), is exactly of the kind required by Postulate B. As for Postulate~A, it is
false iff for every infinite $S$ that is pairwise SLR and a subset of a history and for every product function
on $S$, there is a function $F$, as defined in Equation~(\ref{funkcjaF}). Clearly, in $M_2$ there are plenty of
infinite sets that are pairwise SLR and subsets of a history. Yet, as long as such a set does not contain a
choice point, the construction of a function satisfying the conditions on $F$ is straightforward and we leave it
to the reader. The only case requiring some attention is if a set described above contains a choice point.
Observe that the set of choice points in $M_2$ is $S' = \{\langle 1, n\rangle : n \in \mathbb{N} \}$. Thus, we
need to say what value $F$ takes on elements of $S'$:\footnote{We can
  write this since there is one-to-one correspondence between $Hist$
  and $\mathbb{F}$ in $M_2$.}

$F(\langle 1, n \rangle) := \langle g , \langle 3/2, n, i\rangle \rangle$,\\ where $i = 0,1$ and $g$ is some
function from $\mathbb{F}$ such that $g(n) = i$.  It is easy to see that this function satisfies conditions on
$F$ (\ref{funkcjaF}). Thus, since for every infinite $S$ that is pairwise SLR and a subset of a history and for
every product function on $S$, there is a sought-for function $F$, Postulate~A is false in $M_2$. Finally, as
shown in \cite{muller06:_funny}, there is no funny business in $M_2$.

\bigskip
The upshot of this section is this: generally, there are exactly two ways of producing INFFB, where there is no
case of FINFB: by Postulate~A and by Postulate~B. In the next section, we will investigate if these postulates
can be true in MBS.

\section{Funny business in MBS}\label{FBMBS}\label{sec5}

In this section we ask under what conditions there could be cases of
INFFB in MBS. Since (as we have seen) there are generally two ways of
producing INFFB, namely by our Postulates~A and~B, we also ask what
these postulates amount to in  MBS's.

We begin with showing a simple fact, namely that under some
conditions, there is NO INFFB in MBS. Imagine an infinite, pairwise
SLR set of choice points.  If you can divide the set into finite
``chunks'' separated by a minimal distance, the set does not give rise
to infinitary funny business.

\begin{condition}\label{posC}
  There exists a real number $\delta \in \reals$ such that for any
  infinite, pairwise SLR set of choice points S and for any $x \in \reals^4$, if $x$
  is above only a finite number of points of $S_M$, then so is
  $\langle x^0 + \delta, x^1, x^2, x^3 \rangle \in \reals^4$.
\end{condition}

Notice that truth of condition \ref{posC} in a given model implies, for example, that there are no convergent
sequences in any pairwise SLR set of choice points S in that model.

\begin{fact} Suppose NOFINFB is true in an MBS ${\mathcal W}$. If
  condition \ref{posC} is also true in W, then no $\langle S, f
  \rangle$ such that $S$ is an infinite, pairwise SLR set of choice
  points and $f \in \prod\limits_{e \in S} \Pi_e$ gives rise to INFFB.
\label{fakcik}
\end{fact}

\begin{proof}
  Suppose $S$ is an infinite, pairwise SLR set of choice points, $f
  \in \prod\limits_{e \in S} \Pi_e$ and $\langle S, f \rangle$ gives
  rise to INFFB (*). We will obtain a contradiction by showing that
  there is a history $h_L \in \bigcap\limits_{e \in S} f(e)$, and we
  will arrive at this history by constructing a certain denumerable
  chain $L = \{ l_0, l_1...\}$.

  Take a point $[e_0 \sigma] \in S$\footnote{We hope it
is clear from the context that $e_0 \in \reals^4$ and $\sigma \in \Sigma$. For clarity, in similar cases (when
the point from $\reals^4$ has a subscript) we do not want to write the scenario as a subscript, too.}. Put $l_0
:= [e_0 \sigma]$ and
  $S_0 := \{ [e_0 \sigma] \}$. Since $e_0$ is above only a finite
  number of points of $S_M$ (one, to be exact), we apply condition
  \ref{posC} and put $x_1 := \langle e_0^0 + \delta, e^1. e^2, e^3
  \rangle$. Let $S_1 = (S / S_0) \cap \{ [x_\sigma] : x <_M x_1$ and
  $\sigma$ is a scenario $\}$. $S_1$ is finite. Therefore, by
  NOFINFB, both $\bigcap\limits_{e \in S_0} f(e)$ and
  $\bigcap\limits_{e \in S_1} f(e)$ are nonempty; so, again by
  NOFINFB, $\bigcap\limits_{e \in S_0 \cup S_1} f(e) \neq
  \emptyset$. We take a history $h_{\sigma_1} \in \bigcap\limits_{e
    \in S_0 \cup S_1} f(e)$ and put $l_1 := [x_1 \sigma_1]$. It is
  evident from the construction that $l_0 < l_1$.

  Suppose $l_k = [x_k \sigma_k]$. Since $x_k$ is above only a finite
  number of points of $S_M$ (they all belong to the set
  $\bigcup\limits_{i=0}^k S_i$, which is a sum of a finite number of
  finite sets), we apply condition \ref{posC} and put $x_{k+1} :=
  \langle x_k^0 + \delta, e^1. e^2, e^3 \rangle$. Let $S_{k+1} = (S /
  \bigcup\limits_{i=0}^k S_i) \cap \{ [x_\sigma] : x <_M x_{k+1}$ and
  $\sigma$ is a scenario $\}$. $S_{k+1}$ is finite. Then
  $\bigcap\limits_{e \in \bigcup\limits_{i=0}^k S_i} f(e) \neq
  \emptyset$ (because e.g.  $h_{\sigma_k}$ belongs to it) and
  $\bigcap\limits_{e \in S_{k+1}} f(e) \neq \emptyset$ (by NOFINFB).
  Then, again by NOFINFB, $\bigcap\limits_{e \in
    \bigcup\limits_{i=0}^{k+1} S_i} f(e) \neq \emptyset$. We take a
  history $h_{\sigma_{k+1}} \in \bigcap\limits_{e \in
    \bigcup\limits_{i=0}^{k+1} S_i} f(e)$ and put $l_{k+1} := [x_{k+1}
  \sigma_{k+1}]$. It is evident from the construction that $l_k <
  l_{k+1}$.

  We have described the procedure for creating the chain $L$. By laws
  of BST, there has to be a history $h_L$ containing the chain. We
  claim that $h_L \in \bigcap\limits_{e \in S} f(e)$. Suppose, to the
  contrary, that $h_L \notin \bigcap\limits_{e \in S} f(e)$. So
  $\exists e \in S: h_L \notin f(e)$(**). By our construction, there
  has to be a natural $n$ such that $e \in S_n$. Since $h_L \equiv_e
  h_{\sigma_n}$ and $h_{\sigma_n} \in f(e)$, we get that $h_L \in
  f(e)$, which contradicts (**). We know, then, that $h_L \in
  \bigcap\limits_{e \in S} f(e)$, which contradicts (*).
\end{proof}

\bigskip We know now that under some conditions there is NO INFFB in
an MBS. Since one way of introducing INFFB is via Postulate~A, let us
investigate what this postulate amounts to in an MBS.  It will turn
out that we can introduce a certain kind of funny business (``epsilon
funny business'', labeled $\epsilon FB$), present whenever Postulate A
is true. This will give us some details about the situations in which
funny business can arise in MBS'.

\begin{definition}
Let $S \subseteq W$. We will say that $S_M = \{z \in \reals^4 : \exists \sigma \in \Sigma : [z_\sigma] \in S \}$
is a reduced set derived from $S$.
\end{definition}

In other words, the reduced set of a subset $S$ of $W$ is what we get after projecting $S$ on $\reals^4$; it is
the set of spatiotemporal locations occupied by members of $S$.

\begin{definition}
  We say that in an MBS model ${\mathcal W}$ an infinite pairwise SLR
  set $S$, whose all elements have the form $[z_\sigma]$ for a given $\sigma \in \Sigma$, and a
  product function $f$ constitute $\epsilon$FB iff

  \[\exists e^* \in S_M\; \forall \epsilon(e^*) \bigcap_{e' \in
    \epsilon(e^*)} f([e'\sigma])= \emptyset ,
\]
where $S_M$ is the reduced set derived from $S$, $\epsilon(e^*) = \{e \in S_M |d(e, e^*) < \delta \}$ for some
$\delta \in \Re$ and $d()$ is the Euclidean distance.
\end{definition}

\begin{theorem}\label{eps1}
  If in an MBS model ${\mathcal W}$ there is (1) an infinite pairwise
  SLR set $S= \{[z_{\sigma} ]: z \!\in Z \subset \! \Re^4\}$ and (2) a
  product function $f$ such that $S$ and $f$ constitute $\epsilon$FB,
  then Postulate A is true in ${\mathcal W}$.
\end{theorem}
\begin{proof} To prove Postulate A, take $S$, $f$ and some $e^*$,
  whose existence is postulated by $\epsilon$FB . As a sidenote,
  observe that, since No FIN FB is assumed to be true, it follows from
  $\forall \epsilon(e^*) \bigcap\limits_{e' \in \epsilon(e^*)}
  f([e'_\sigma])= \emptyset$ that every neighborhood $\epsilon(e^*)$
  comprises infinitely many elements of $S_M$. This is possible only
  if $e^*$ is a point of convergence of a sequence in $S_M$ (a special
  case is dense $S_M$).

  We need to show that for every triple $\langle \gamma, x, \alpha
  \rangle$ ($\gamma, \alpha \in \Sigma, x \in \Re^4$) the following
  formula is true:

\begin{equation}
\begin{split}
  [x_\alpha] \neq [x_\gamma] \vee [e^*_\sigma] \neq [e^*_\alpha]
  \vee x\not> e^* \vee \gamma \not\in f([e^*_\sigma]) \vee \\
  \exists [e'_\sigma] \in S (\gamma \not\in f([e'_\sigma]) \wedge
  \neg(x \text{ SLR}_M e')).
\end{split}
\label{eq:5dis}
\end{equation}

\noindent The formula is the disjunction $D_1 \vee D_2 \vee D_3 \vee
D_4 \vee D_5$.  For a triple $\langle \gamma, x, \alpha \rangle$ we
consider now two cases. It may be that for that triple $D_1 \vee D_2
\vee D_3 \vee D_4$ is true, and hence, obviously, the entire
disjunction (\ref{eq:5dis}) is true as well. In the other case, $D_1
\vee D_2 \vee D_3 \vee D_4$ is not true for the triple, i.e., the
conjunction $\neg D_1 \wedge \neg D_2 \wedge \neg D_3 \wedge \neg D_4$
obtains.  For Postulate A to be true, in this case the triple must
make $D_5$ true, i.e.,

\begin{equation}
\exists [e'_\sigma] \in S\; (\gamma \not\in f([e'_\sigma]) \wedge
  \neg(x \text{ SLR}_M e')).
\label{eq:1}
\end{equation}

\noindent We will now show this. Since by $\neg D_3$: $x > e^*$, we may define a nonempty subset $B(x)$ of
$S_M$, whose elements are in the backward light cone of $x$, i.e., $B(x) := \{z \in S_M| z <_M x \}$.  Observe
now that for all $e$ from $S_M$ (a) $e \in B(x)$ iff $\neg(e \text{ SLR}_M\; x)$. The implication to the right
is obvious; to see that the reverse implication holds as well, note that it holds if $e' < x$. And $e' > x$ is
impossible, since it implies (as $x > e^*$) that for some $e \in S_M$ $e' > e$, which contradicts that $S_M$ is
pairwise SLR.

Now by (a), Formula~\ref{eq:1} is equivalent to:

\[
\exists [e'_\sigma] \in S\; (\gamma \not\in f([e'_\sigma]) \wedge e' \in B(x)),
\]
which is equivalent to:
\[
\neg \forall [e'_\sigma] \in S\; (e' \in B(x) \rightarrow \gamma \in f([e'_\sigma])),
\]
and further equivalent to:

\begin{equation}
  \gamma \not\in \bigcap_{e' \in B(x)} f([e'_\sigma]). \label{eq:2}
\end{equation}

\noindent From the construction of $B(x)$ and $\epsilon$FB : \[\exists
\epsilon(e^*) \; \epsilon(e^*) \subseteq B(x) \;\wedge \; \bigcap_{e'
  \in \epsilon(e^*)} f([e'_\sigma]) = \emptyset.\] It follows that
$\bigcap\limits_{e' \in B(x)} f([e'_\sigma]) = \emptyset$. So it cannot be that $\gamma \in \bigcap\limits_{e'
\in B(x)} f([e'_\sigma])$, and thus the formula~\ref{eq:2} and hence $D_5$ must be true.

\end{proof}

\begin{theorem}\label{eps2}
  If Postulate A is true in an MBS model ${\mathcal W}$, then there is in ${\mathcal W}$ an infinite pairwise
SLR set $S$, whose all elements have the form $[z_\sigma]$ for a given $\sigma \in \Sigma$, and a product
function $f$ such that $S$ and $f$ constitute $\epsilon$FB. \label{the:AtoINF*}
\end{theorem}

\noindent \begin{proof} Let Postulate A be true in ${\mathcal W}$ due
  to $S$, $f$, and $e^*$.  For \emph{reductio}, assume that $S$ and
  $f$ do not constitute $\epsilon$FB. We get that $\exists
  \epsilon(e^*) \bigcap\limits_{e' \in \epsilon(e^*)} f([e_\sigma])
  \neq \emptyset$. We will construct a triple $x, \gamma, \alpha$ that
  contradicts Postulate A. Take the $\gamma \in \bigcap\limits_{e'
    \in \epsilon(e^*)} f([e_\sigma])$ for some $\epsilon(e^*)$ and $x
  >_M e^*$ such that $B(x) \subseteq \epsilon(e^*)$. (It is always possible
  to find such an $x$ since we are in an MBS and thus are searching in $\reals^4$.) Accordingly,
  $\gamma' \in \bigcap\limits_{e \in B(x)} f([e_\sigma ])$. Because
  $\forall e'\; ( e' \in S / B(x) \rightarrow x\; SLR_M\; e')$, we
  arrive at a contradiction with $D_5$. Since clearly $e^* \in
  \epsilon(e^*)$, we have $\gamma \in f([e^*_\sigma])$, which
  contradicts $D_4$.  $x >_M e^*$ contradicts $D_3$. To obtain a
  contradiction with $D_1$, let $\alpha =
  \gamma$. Since $\gamma' \in f([e^*_\sigma])$, $[e^*_\sigma] =
  [e^*_\gamma']$, and hence $[e^*_\sigma] = [e^*_\alpha]$, which
  contradicts $D_2$. Thus, the triple $x, \alpha, \gamma'$
  contradicts the disjunction $D_1 \vee D_2 \vee D_3 \vee D_4 \vee
  D_5$, and hence Postulate A.
\end{proof}

\bigskip Theorems \ref{eps1} and \ref{eps2} give us the following fact
about the equivalence of Postulate A and epsilon funny business:

\begin{fact}
  Let ${\mathcal W}$ be an MBS model. The following conditions are
  equivalent:

(1) Postulate A is true in ${\mathcal W}$

(2) There is in ${\mathcal W}$ an infinite pairwise SLR set $S= \{[z_\sigma ]: z
  \!\in\! S_M \subseteq \Re^4\}$ and a product function $f$ such that
  $S$ and $f$ constitute $\epsilon$FB.
\label{epsi-vs-A}
\end{fact}

Note the consequences of this fact. In order Postulate~A be true in an MBS, there must be in the reduced set
$S_M$ such an $e^*$ that $\bigcap\limits_{e' \in \epsilon(e^*)} f([e'_\sigma])= \emptyset$, no matter how small
the diameter of $\epsilon(e^*)$ is. The intersection cannot be empty if it is possible to have an
$\epsilon(e^*)$ to which only $e^*$ belongs.  Also, on the supposition that NO FINFB is true, the intersection
cannot be empty if some $\epsilon(e^*)$ contain only a finite subset of $S_M$.  Hence, in MBS Postulate~A can be
true provided that $S_M$ contain a convergent sequence, together with its point of convergence.

\bigskip

We will finally check whether Postulate~B is false in MBS'. To this
end, it would be enough to prove that in an MBS, if $\neg$~Postulate~A and
NO FINFB were true, then NO INFFB was true as well. Given the
Theorem~(\ref{tt2}), this would imply that Postulate~B was false in MBS'.

To begin our exploration, let us ask what form a pairwise SLR set $S$,
which is a subset of a history, should take to give rise to funny
business, INFFB or FINFB, in an MBS.  Assume thus that $S$ is a subset
of $W$, all elements of which have the form $[e_\sigma]$ for a given
$\sigma \in \Sigma$. Clearly, $S$ is a subset of a history. Consider a
product function $f$ on $S$, and assume that $\langle S, f\rangle$ is
a case of funny business, FINFB or INFFB. Thus,\[ \bigcap_{[e_\sigma]
  \in S} f([e_\sigma]) = \emptyset.\] Pick now some $a^*= \langle
a_0^*, a_1^*, a_2^*, a_3^* \rangle \in S_M$ and consider elements of
$\Re^4$ located `vertically' above $a^*$, i.e., the set $L := \{x\mid
x = \langle x_0, a_1^*, a_2^*, a_3^* \rangle \wedge x_0 \geqslant
a^*_0\}$.

Our guiding idea is this: we want to see what has to happen if
$\langle S, f \rangle$ is to be a case of funny business, FINFB or
INFFB.  Usually there exist `safe' subsets of $S$ such that there is a
history passing through all the elementary possibilities designated by
the function $f$ at the points from the subsets. Suppose that it is
not the case regarding the whole set $S$ (meaning that we have an
example of funny business). What, then, with the `in-between' section
- the sets larger then the obviously `safe' ones and smaller than the
whole $S$? Does there have to be a biggest `safe' set, or a smaller
`unsafe' set? Let us put these ideas down formally. For $x \in L$, we
will say that $x$ is good iff for $S_x := \{ [e_\sigma] \in S\mid e
\leqslant_M x\}$: $\bigcap\limits_{[e_\sigma] \in S_x} f([e_\sigma])
\neq \emptyset$. If $x$ is not good, we will say it is bad.

Clearly, $a^*$ is good, and if $x, y \in L$, $x <_M y$, and $y$ is
good, then $x$ is good as well. Hence there are initial non-empty
segments of $L$, whose all elements are good. We call such initial
segments ``good''. The question now is: what maximal good initial
segments of $L$ are possible? (Since the sum of a set of good segments
is a good segment, for any chain a maximal good initial segment will
exist.) There are two possibilities:

(1) The maximal good initial segment of $L$ is not upper bounded, and hence is identical to $L$. By an argument
similar to that given at the end of our proof of Fact~\ref{fakcik}, there is a history $h_L \in
\bigcap\limits_{[e_\sigma] \in S} f([e_\sigma])$; hence $\langle
S, f\rangle$ is not a case of funny business, i.e., we have neither FINFB nor INFFB.\\
(2) The maximal good initial segment of $L$ is upper bounded, so by
properties of $\Re^4$ and our construction, it has a supremum; dub
it $x^*$. Is then $x^*$  good or bad? Again, there are two options:

\indent \indent (2a) $x^*$ is bad. We already know that for any $x < x^*, x \in L$ is good. Hence, funny
business is located, so to speak, on the backward light cone of $x^*$, i.e., on $B_{\mid x^*} = \{x \in \Re^4
\mid x <_M x^* \wedge D^2_M(x, x^*) = 0\}$ (where $D^2_M$ is the Lorentz interval).  In other words, defining $S' =
\{[e_\sigma] \in S \mid e \in  B_{\mid x^*}\}$, we have it that $\langle S' f\rangle$ is a case of funny
business, FINFB or INFFB.

\indent \indent (2b) $x^*$ is good. This means that any $x > x^*, x
\in L$ is bad. Hence, on any outer lining (however thin)
$B_{x^*}(\delta) = \{x \in \Re^4 \mid x <_M x^*(\delta) \wedge x
\not\leqslant_M x^*\}$, where $x^*(\delta)= \langle x^*_0 + \delta,
x^*_1, x^*_2, x^*_3 \rangle$, of the backward light cone of $x^*$,
there is located a case of funny business.  More precisely, defining
$S(\delta) = \{[e_\sigma] \in S \mid e \in B_{x^*}(\delta)\}$, we have
it that for any $\delta$, $\langle S(\delta), f\rangle$ is a case of
funny business, FINFB or INFFB.

\medskip Thus, the result of this exploration is that if a pairwise
SLR set $S$ (where $S \subseteq h$ for some history) gives rise to
funny
business in an MBS, then there is $x^* \in \Re^4$ and one of two conditions  obtains:\\
(1) on the backward light cone of $x^*$ there is the reduced set
$S'_M$ of $S'$ such that $S' \subseteq S$ and $S'$ gives rise to funny
business,
or \\
(2) on any outer lining of the backward light cone of $x^*$ there is
the reduced set $S_M(\delta) \in \Re^4$ of $S(\delta)$ such that
$S(\delta) \subseteq S$ and $S(\delta)$ gives rise to funny business.
To put it more precisely,

\begin{fact}
  If a pairwise SLR set $S$, which is a subset of a history gives rise
  to funny business (FINFB or INFFB) in an MBS, then there is $x^* \in
  \Re^4$ such that either (1) there is a set $S'\subseteq S$ giving rise to
  funny business, whose reduced set $S'_M$ is a subset of $B_{\mid
    x^*} = \{x \in \Re^4 \mid x <_M x^* \wedge D^2_M(x, x^*) = 0\}$, or
  (2) there is a family of sets $S(\delta)\subseteq S$, $\delta > 0$,
  each $S(\delta)$ giving rise to funny business, and such that the
  reduced set $S_M(\delta)$ of $S(\delta)$ is a subset of
  $B_{x^*}(\delta) = \{x \in \Re^4 \mid x <_M x^*(\delta) \;\wedge\; x
  \not\leqslant_M x^*\}$, where $x^*(\delta)= \langle x^*_0 + \delta,
  x^*_1, x^*_2, x^*_3 \rangle$.
\label{fakcisko}
\end{fact}
Note that $S$ can be different from $S'$ as well as from any
$S(\delta)$.

Let us observe the first consequence of this exploration: in MBS-like
models produced on $\Re^2$, a pairwise SLR set which is a subset of a
history cannot give rise to INFFB if $\neg$ Postulate~A and NOFINFB
are true.\footnote{Strictly speaking, we defined MBS models as based
  on $\Re^4$, i.e., elements of our models are equivalence classes
  $[x_\sigma]$, where $x \in \Re^4$. There is no obstacle, however, to
  consider MBS-like models based on $\Re^n$ of different
  dimensionality.} In the two dimensional case, the backward light
cone $B_{\mid x^*}$ is reduced to two straight lines, call them ``left'' and ``right''. Since $S =
\{[e_\sigma]\mid e \in S_M\}$ ($\sigma \in \Sigma$) is assumed to be pairwise SLR, $S_M$ could have at most two
elements located on $B_{\mid x^*}$ (one of the left line, the other on the right line).  Thus, in
two-dimensional case, if NOFINFB is true, there can be no INFFB located on the backward light cone of $x$.  Can
there be INFFB located on outer linings $B_{x^*}(\delta)$? Consider outer linings of one line, say left one.
Such outer linings can be written as $B_L(\delta) = \{\langle y_0, y_1\rangle \in B_{x^*}(\delta)\mid y_1 <
x^*_1\}$. Note that, for $e \in S_M$, if $e \in B_L(\delta)$, then no $e' \in S_M$ such that $e' <_M e$ can
belong to $B_L(\delta)$. In a similar vein, no $e' \in S_M$ such that $e' >_M a^*$ can belong to $B_L(\delta)$,
where $a^*$ is that element of $S_M$ from which we started the construction of the chain $L$ containing $x^*$.
Thus, $B_L(\delta)$ cannot go infinitely down along the left line, or up to $x$. Hence, given that NOFINFB is
true, in order every $B_L(\delta)$ (i.e., for any $\delta > 0$) generate INFFB, there must be a sequence of
elements of $S_M$ converging to a point from the left line. Moreover, since elements of $S_M$ are SLR$_M$, there
must be exactly one point of convergence located on the left line. The same is true about right outer linings,
$B_R(\delta)$. That is, for INFFB to obtain, a sequence in $S_M$ must converge to a point $e_L$ located on the
left line or a sequence in $S_M$ must converge to a point $e_R$ located on the right line. Accordingly, we
should consider three cases: (1) in $S_M$ there is only a sequence converging to a point $e_L$ on the left line,
(2) in $S_M$ there is only a sequence converging to a point $e_R$ on the right line, and (3) in $S_M$ there are
two sequences, one converging to a certain $e_L$ on the left line and the other converging to a certain $e_R$ on
the right line.  In each case we need to construct an outer lining $B_{x^*}(\delta^*)$ that does not generate
INFFB. We will produce the required construction for case (3) only, as it is more complicated, and the reader
will surely know how to transform it into arguments appropriate for the remaining two cases.  Assume then case
(3); it might happen that $e_L$ (or $e_R$ or both) does not belong to $S_M$.  However, if $S_M \cup \{e_L, e_R\}
:= S_M'$ does not generate INFFB, $S_M$ does not generate it either.  Consider thus the `extended' set $S_M'$.
Assume $\neg$ Postulate~A, which (as we have seen in Fact~\ref{epsi-vs-A}), is equivalent to NO $\epsilon$FB.
Apply then NO $\epsilon$FB to $e_L$ and $e_R$, obtaining $\epsilon (e_L)$ and $\epsilon(e_R)$, estimated, resp.,
by diameters $\delta_L$ and $\delta_R$. Put $\delta^* = min\{\delta_L, \delta_R \}$ and consider the outer
lining $B_{x^*}(\delta^*)$. By our construction, the elements of $S_M$ that belong to $B_{x^*}(\delta^*)$, but
neither to $\epsilon(e_L)$ nor to $\epsilon(e_R)$ constitute a finite set. By NO $\epsilon$FB we have, for any
product function $f$ on $S' = S \cup \{[e_L \sigma], [e_R \sigma]\}$: $\bigcap\limits_{e \in
  \epsilon(e_L)}\; f([e_\sigma]) \neq \emptyset$ and
$\bigcap\limits_{e \in \epsilon(e_R)}\; f([e_\sigma]) \neq \emptyset$.
Combining all these observations together, and assuming NOFINFB, we
have an outer lining $B_{x^*}(\delta^*)$ that does not generate INFFB,
i.e.

\noindent
$\bigcap\limits_{e \in S_M \cap B_{x^*}(\delta^*)}\; f([e_\sigma]) \neq \emptyset$,
which contradicts the result that every outer lining $B_{x^*}(\delta)$
generates INFFB.

\bigskip One might hope that this result, i.e., in 2-dimensional
MBS-like model, a pairwise SLR set $S$, which is a subset of a history
cannot give rise to INFFB if NOFINFB and $\neg$~Postulate~A are true,
carries over to the real, i.e, 4-dimensional MBS. This hope is however
shattered by a construction, which in essence consists in wrapping an
$M_2$ structure around a backward light cone of some $x^*$.

Consider an MBS whose set $\Sigma$ of history labels is the set of all
functions $g:\mathbb{N} \rightarrow \{0,1\}$ such that for only
finitely many $n \in \mathbb{N}$, $g(n) = 0$. Our construction of
Section~\ref{shape} guarantees that this is indeed an MBS, provided
that the topological Postulate~\ref{p1} is true, which we will check
in a due course.  Consider $x^* = \langle 0,0,0,0\rangle \in \Re^4$,
its backward light cone $B_{\mid x^*}$, and a sequence of `angles'
$\varphi_n = \pi (2^n - 1)/2^n$ ($n = 1, 2, \ldots$). Let $S_M =
\{\langle -n, n \cos(\varphi_n), n \sin(\varphi_n), 0 \rangle \mid n =
1, 2, \ldots \}$. The (Euclidean) distance between any two elements of
$S_M$ is at least $\sqrt{2}$.  Also, $S_M$ is pairwise SLR$_M$ and
$S_M \subseteq B_{\mid x^*}$.  $e_n \in S_M$ belongs to appropriate
sets of splitting points, according to this rule:

\begin{equation*} \text{for }g, g' \in \Sigma\text{, }e_n \in C_{g
    g'}\text{ iff }g(n) \neq g'(n).
\end{equation*}

We can assign the same history label to all elements of $S_M$, for
instance $g^*$ such that for all $n$, $g^*(n) = 1$. Consider a product
function $f$ on $S = \{[e_n g^*]\mid e_n \in S_M\}$ defined as
\begin{equation*} f([e_n g^*]) := \{g \in \Sigma \mid g(n) = 0\}.
\end{equation*}
(Compare our discussion of elementary possibilities in $M_2$ on page
\pageref{hisM2}.) Since among history labels there is no function $g$
that yields infinitely many zeros, $\langle S, f\rangle$ is a case of
INFFB. On the other hand, $S$ does not give rise to FINFB, since there
are history labels yielding arbitrarily large finite number of zeros.
Also, since there is a minimal distance $\sqrt{2}$ between elements of
$S_M$, one can associate with every $e_n \in S_M$ an $\epsilon(e_n) =
\{e \in S_M |d(e, e_n) < \sqrt{2}/2\}$ to which only the singleton
$\{e_n\}$ belongs. The existence of such epsilons means that
NO~$\epsilon$FB is true in our model, and hence, by
Fact~\ref{epsi-vs-A}, $\neg$~Postulate~A is true as well. This means
that Postulate~B must be true in the model.\footnote{A reader might
  want to directly see an odd set $X$ of which Postulate~B is true. To
  this end, from each $e_n \in S_M$ go up a bit (say, by $1/2$ on the
  temporal axis) along the backward light cone of $x^*$ and with the
  resulting point $x_n \in \Re^4$ associate a label $g$ such that g(n)
  = 0. The totality of $[x_n g]$ ($n \in \naturals$) is the sought-for
  $X$.}

Returning to the topological postulate, the only chains in $\Re^4$ that might falsify it are those that contain
$x^*$. As an example, consider the vertical chain $L$ of elements below $x^*$, including $x^*$. Since histories
are labeled by elements of the set $\Sigma$, we might write, for $z \in L$, $\Sigma_g(z) := \{ g' \in \Sigma
\mid [z_g] = [z_{g'}] \}$, like in Definition~\ref{Sigma}.

We have now, for every $y \neq x^* \in L$: $\Sigma_g(y) = \Sigma$,
whereas $\Sigma_g(x^*) = \{g\}$. However odd it looks like, it means
that our topological postulate is satisfied, and hence the described
model is indeed an MBS. The oddity is similar to a feature of $M_2$:
no matter which history label is associated with $x^*$, the resulting
event in BST is the smallest upper bound of some elements of $S$ and
it cannot be SLR to any remaining element of $S$. The second trick of
our MBS is that elements of $S_M$ escape steadily to the past, which
ensures that there is always a finite distance between them. A natural
question is whether $S_M$ bounded on temporal coordinates can yield
INFFB if NO FINFB and $\neg$~Postulate~A are true; we leave it as an
open problem.

\bigskip
To sum up our investigations of MBS', we have the following:

\noindent
(1) If NO FINFB and condition \ref{posC} are true, then no infinite
and pairwise SLR set $S$ gives rise to INFFB.

\noindent
(2) If in an MBS there is funny business (FINFB or INFFB) produced by
a pairwise SLR subset $S$ of a history, then there is some $x \in
\Re^4$ such that either (1) a set $S'\subseteq S$, which gives rise to funny
business and whose reduced set $S'_M$ is located on the backward light
cone of $x$, or (2) any outer lining of backward light cone of $x$
comprises the reduced set of a set $S(\delta) \subseteq S$ which
generates funny business.

\noindent
(3) In order for Postulate~A to be true, given that NOFINFB obtains,
there must be a set of choice points whose reduced set contains a
converging sequence together with its point of convergence.

\noindent
(4) Postulate~B cannot be true in 2-dimensional MBS-like models if
$\neg$~Postulate~A and NOFINFB are true. However, there are
4-dimensional MBS models in which Postulate~B is true, even although
$\neg$~Postulate~A and NOFINFB are true. It is an open question
whether for such cases to arise one needs a set $S_M$ which is not
bounded on the temporal coordinate.

\section{Conclusion and open problems}\label{concl}
In this paper we have introduced the notion of a Minkowskian Branching
Structure, based on M\"{u}ller (\citeyear{muller_nato}). In the second
part of the paper we have shown some results concerning finitary and
infinitary funny business. There are exactly two ways of generating
INFFB which does not involve FINFB: via Postulate A or via Postulate
B.  On the other hand, the falsity of both  postulates in a model
with space-time points and NOFINFB entails that there is NOINFFB in
this model. In the third part of the paper we explored under what
conditions there could be INFFB in MBS.  We first observed that if in
an MBS model  choice points are `nicely' distributed in
$\Re^4$, so that condition \ref{posC} is true, and NOFINFB is true,
then there is no case of INFFB in that model. Also, we have seen that
in MBS' a set $S$ responsible for funny business must be quite
particularly located, namely if there is funny business produced by a
pairwise SLR subset $S$ of a history, then there is some $x \in \Re^4$
such that either (1) there is a set $S'\subseteq S$, which gives rise to funny
business and whose reduced set $S'_M$ is located on the backward light
cone of $x$, or (2) any outer lining of the backward light cone of $x$
comprises the reduced set of a set $S(\delta) \subseteq S$ which
generates funny business.  Our next finding is that Postulate~A can be
true in MBS' (and generate INFFB) only provided that there is a convergent
sequence in reduced set $S_M$ of $S$ which gives rise to INFFB.
Finally, we have shown an MBS model in which Postulate~B is true, yet
Postulate A and FINFB do not hold; the set $X$ of which Postulate~B
speaks is not bounded on the time coordinate. We conjecture that for
Postulate~B to be true in an MBS, `its' set $X$ must have that
feature. The moral of our findings is that, spatio-temporally
speaking, INFFB without FINFB is possible; however, since this
phenomenon requires either convergent (in $\Re^4$) sequences of choice
points, or choice points escaping to infinity on the time coordinate,
INFFB without FINFB does not seem to be physically possible.

\newpage
\section{Appendix}
\subsection{Splitting points and choice points}\label{spcp}
Since it purports to establish that ``For histories $h_\sigma$,
$h_\eta$ $\subset B$ the set $C_{\sigma,\eta}$ is the set of choice
points'', Lemma 4 in M\"{u}ller seems to require reformulation. A
splitting point, as a member of $\reals^4$, is not a member of $B$,
and thus is not a choice point.

An obvious move would be to observe that every splitting point $x$ for
scenarios $\sigma$ and $\eta$ in a sense ``generates'' a choice point
for histories $h_\sigma$ and $h_\eta$. That is, if $x \in
C_{\sigma\eta}$ then $[x_\sigma]$ is maximal in $h_\sigma \cap
h_\eta$.

What might not be as evident is that, since we have dropped the
requirement of finitude of every $C_{\sigma\eta}$, the converse is not
true: in some cases there are choice points which are not
``generated'' in the above way by any splitting points. We will now
try to persuade the reader that this is indeed the case. The idea is
to use sequences of generated splitting points convergent to the same
point. The argument is simple in $\reals^2$ as we need only two
sequences, but gets more complicated as the number of dimensions
increases. (For convenience, in the below argument we use symbols
``$>_S$'' and ``$>_M$'' defined in the natural way basing on
respectively ``$\leqslant_S$'' and ``$\leqslant_M$''.)

\begin{definition}
1. $SC_{\sigma\eta}:=\{[c_\sigma] \mid c \in C_{\sigma\eta}\}$
\begin{eqnarray*}2. \textbf{C}_{\sigma\eta}:=\{[x_\gamma]:&(1)&
[x_\gamma] \in h_\sigma \cap h_\eta\mbox{~and}\\ &(2)& \forall z \in \reals^4 \forall \alpha \in \Sigma
([z_\alpha]
>_S [x_\gamma] \Rightarrow [z_\alpha] \notin h_\sigma \cap h_\eta)\
\end{eqnarray*}
\end{definition}

``$SC_{\sigma\eta}$'' is to be read as ``The set of generated choice
points for histories $h_\sigma$ and $h_\eta$''.

``$\textbf{C}_{\sigma\eta}$'' is to be read as ``The set of choice
points for histories $h_\sigma$ and $h_\eta$''.

It is of course irrelevant whether we choose $\sigma$ or $\eta$ in
square brackets in the definition of the set of generated choice
points, since if $c \in C_{\sigma\eta}$ then $c_\sigma \equiv_S
c_\eta$ and thus $[c_\sigma] = [c_\eta]$.

\begin{theorem}\label{LW1}{For some $C_{\sigma\eta}$, $SC_{\sigma\eta}
\varsubsetneq \textbf{C}_{\sigma\eta}$.}
\end{theorem}

\textit{Proof sketch.} Again, by fixing two spatial dimensions we will
restrict ourselves to $\reals^2$. Let $x=(0,0)$. Let $C_1=\{(0,1/n)|n
\in \mathbb{N} \backslash \{0\}\}$ and $C_2=\{(0,-1/n)|n \in
\mathbb{N} \backslash \{0\}\}$. Let $C_{\sigma\eta} = C_1 \cup C_2$.
As $x \notin C_{\sigma\eta}$, it is evident that $[x_\sigma] \notin
SC_{\sigma\eta}$. We will show that $[x_\sigma] \in
\textbf{C}_{\sigma\eta}$, thus proving the theorem.

We have to show that $[x_\sigma]$ meets conditions $(1)$ and $(2)$
from the above definition. As for $(1)$, $ \forall c \in
C_{\sigma\eta} \mbox{ x SLR c }$, so $x \in R_{\sigma\eta}$. It
follows that $x_\sigma \equiv_S x_\eta$ and finally (as it is obvious
that $[x_\sigma] \in h_\sigma$) that $[x_\sigma] \in h_\sigma \cap
h_\eta$.

Now for (2). Consider $[z_\alpha]$ such that (a) $[z_\alpha] >_S
[x_\sigma]$. By definition of $>_S$, $z >_M x$ and $x_\alpha =
x_\sigma$. Let $z=(z_0,z_1)$ (the first coordinate is temporal). We
distinguish two cases: either the spatial coordinate $z_1$ is equal to
$0$ or it's something else.

If $z=(z_0,0)$, take $k \in \reals$, $k<z_0$ such that $(0,k) \in
C_{\sigma\eta}$ (such $k$ exists since $C_1$ converges to $(0,0)$).
(*) Since $D^2_M(z,(0,k)) = k - z_1 <0$, it follows that $x >_M (0,k)
\in C_{\sigma\eta}$.

On the other hand, if $z_1 \neq 0$, consider v defined as follows:
\[
v :=\left\{\begin{array}{lll}1 & \textrm{if $z_1 \geq 1$}\\ z_1 & \textrm{if $z_1 \in (0,1) \cup (-1,0)$}\\-1 &
\textrm{if $z_1 \leqslant -1$}\end{array}\right.
\]

We choose $(0,k) \in C_{\sigma\eta}$ such that $0<k \leqslant v$ (if $v$ is positive) or $v \leqslant k<0$ (if
$v$ is negative). It is always possible to find such a point since both $C_1$ and $C_2$ converge to $(0,0)$. We
have to prove that (b) $z >_M (0,k)$.

From (a) we know that (c) $z >_M (0,0)$. To arrive at (b) it suffices to show that (d) $z >_M (0,v)$. From (c)
it follows that (e) $z_0 \geq z_1$. We have two cases to consider. First, if (f) $z_1 \geq 1$ or $z_1 \leqslant
-1$, $D^2_M(z,(0,v)) = -z_0^2 + (z_1 - 1)^2 = -z_0^2 + z_1^2 + 1 - 2z_1$, which (by (f) and (e)) is below $0$,
which fact is equivalent to (d). Second, if $z_1 \in (0,1) \cup (-1,0)$, $D^2_M(z,(0,v)) = -z_0^2 + (z_1 -
z_1)^2 = -z_0^2$ which is of course negative, so again we arrive at (d).

From (c) and (d) and from the requirement on choosing $(0,k)$ we get the needed result (b).

Since $z >_M (0,k) \in C_{\sigma\eta}$, it is true that $z \notin
R_{\sigma\eta}$ and thus $[z_\alpha] \notin h_\sigma \cap h_\eta$. We
have thus proved that $[x_\sigma]$ fulfills condition (2).

Unfortunately already in $\reals^3$ the construction fails at point (*). To overcome the problem we would have
to use four sequences of splitting points convergent to $(0,0,0)$ (intuitively situated at the arms of the
coordinate system). To deal with the situation in $\reals^4$ we would have to similarly introduce six sequences
convergent to $(0,0,0,0)$. We don't dwell into the details here as the point being made doesn't seem to be
significant enough in proportion to the arduous complexity of the argument.

\begin{conjecture}
For any scenarios $\sigma, \eta \in \Sigma$, the set $\textbf{C}_{\sigma\eta}$ contains exclusively points which
belong to $SC_{\sigma\eta}$ or points $[x_\alpha]$ such that $x$ is a limit of a sequence of points belonging to
$C_{\sigma\eta}$.
\end{conjecture}

\subsection{When the topological postulate is false}\label{imptop}

We will now show a situation in which lemma \ref{l322} does not hold.
The construction resembles the $M_1$ structure from
\cite{muller06:_funny}. By fixing two spatial dimensions we will restrict
ourselves to $\reals^2$, the first coordinate representing time.

As usual, $\Sigma$ is the set of all scenarios of a world $B$. Let $C$ be the set of all splitting points: \[ C
:= \bigcup_{\sigma,\eta \in \Sigma} C_{\sigma\eta}
\] We put
\begin{equation}\label{imptop1}
C := \{ \langle 0, n \rangle | n \in \mathbb{N} \cup \{ 0 \} \}
\end{equation}
The idea is that all splitting points are binary: any scenario passing through a given splitting point can go
either ``left'' or ``right''. Since there are as many splitting points as natural numbers, we can identify
$\Sigma$ with a set of 01-sequences. Another requirement on $\Sigma$ is that it contains only the sequences with
finitely many 0s. Let $G$ be a subset of $\Sigma$ containing only the sequence without any 0s and all sequences
that have all their 0s in the beginning. The elements of $G$ will be labeled as below:
\begin{eqnarray*}\label{imptop2}
\sigma_0 = 1111..... \\ \sigma_1 = 01111.... \\ \sigma_2 = 00111....
\\ \sigma_3 = 00011....
\end{eqnarray*}
Let us next consider a sequence $Z_i^M$ of points in $\reals^2$ such
that for all $i \in \mathbb{N} \; z_i = \langle i-1/2,0 \rangle$. This
way, a given $z_i \in Z_i^M$ is in the Minkowskian sense above all
splitting points $\langle 0, n \rangle | n < i$ and above no other
splitting points.

Consider now a sequence $Z_i$ in $B$, $Z_i = \{ [z_i \sigma_i] | i \in \mathbb{N} \}$\footnote{Again, we hope it
is clear from the context that $z_i \in \reals^4$ and $\sigma_i \in \Sigma$. For clarity, we do not want to
write the scenario as a subscript.}. We will now show that $Z_i$ is a chain. Take any $[z_m \sigma_m],[z_n
\sigma_n] \in Z_i$ such that $m \neq n$. Either $m < n$ or $n < m$; suppose $m < n$ (the other case is
analogous). Since $m < n$, $z_m \leqslant_M z_n$. $z_m \in R_{\sigma_m \sigma_n}$ since it is not above any
splitting points between $\sigma_m$ and $\sigma_n$. Therefore $z_m \sigma_m \equiv_S z_m \sigma_n$, so $[z_m
\sigma_m] \leqslant_S [z_n \sigma_n]$. We have shown that any two elements of $Z_i$ are comparable by
$\leqslant_S$. Therefore, $Z_i$ is a chain in $B$, thus being an upward-directed subset of $B$.

The set of all upward-directed subsets of $B$ meets the requirements
of Kuratowski-Zorn Lemma, since a set-theoretical sum of any chain
subset of it is also an upward-directed subset of $B$ and is an upper
bound of the chain with respect to inclusion. Therefore, there exists
a maximal upward-directed subset of $B$ (a history $h^*$) such that
$Z_i \subseteq h^*$. But lemma \ref{l322} is false with respect to
this history, since for all $\sigma \in \Sigma$, $h^* \neq \{
[x_\sigma]|x \in \reals^2 \}$! Suppose to the contrary, that for a
certain $\sigma \in \Sigma \; h = \{ [x_\sigma] | x \in \reals^2 \}$.
As a member of $\Sigma$, $\sigma$ has to contain a ``$1$'' at some
point $k$ (starting with $0$). Then both $[z_{k+1} \sigma_{k+1}] \in
h^*$ and $[z_{k+1} \sigma] \in h^*$, so $z_{k+1} \in R_{\sigma_k
  \sigma_{k+1}}$. But $C_{\sigma_k \sigma_{k+1}} \ni \langle 0, k
\rangle \leqslant_M z_{k+1}$, so $z_{k+1} \notin R_{\sigma_k
  \sigma_{k+1}}$ and thus we arrive at a contradiction.

We will now show that our topological postulate \ref{p1} is not met in
this situation. Consider a chain $Z := Z_i^M \cup \{ \langle -1,0
\rangle \}$. Note that $\langle -1,0 \rangle = inf(Z)$. Consider next
the chain topology on $\Sigma_{h^*}(\langle -1,0 \rangle)$ (as defined
in the last section) with $Z$ as the original chain. $\{
\Sigma_{h^*}(z_i) \}$ is a centred family of closed sets, but its
intersection is empty as $\Sigma$ does not contain a scenario
corresponding to the sequence comprised of 0s only. Therefore we
arrived at a contradiction with our corollary \ref{c1}, so the
postulate \ref{p1} is not met: the chain topology is \textit{not}
compact.

\section{Equivalences}

We will show that our notion of finitary funny business is equivalent
to the \citet{belnap_nato} notion of generalized primary slr
modal-correlation funny business and our infinitary funny business is
`almost' equivalent to the \citet{muller06:_funny} notion of
combinatorial funny business.

To recall first Belnap's definition, he says that two initial events
$A$ and $B$ and their two elementary possibilities $\Pi_A\langle
h_A\rangle$ and $\Pi_B\langle h_B\rangle$, resp., constitute a case of
generalized primary slr modal-correlation funny business iff $A \; SLR
\; B$ and $\Pi_A\langle h_A\rangle \cap \Pi_B\langle h_B\rangle =
\emptyset$, where $A \subset h_A$ and $B \subset h_B$ (initial event
is an upper bounded subset of a history).

\begin{lemma}
  If $A$ and $B$ and their two elementary possibilities $\Pi_A\langle
  h_A\rangle$ and $\Pi_B\langle h_B\rangle$, resp., constitute a case
  of generalized primary slr modal-correlation funny business, then
  any $\langle S,f\rangle$ such that $A \cup B \subseteq S$ and $f \in
  \prod_{e \in S} \Pi_e$ and $f(e) = \Pi_e\langle h_A\rangle$ if $e
  \in A$ and $f(e) = \Pi_e\langle h_B\rangle$ if $e \in B$, is a case
  of finitary funny business.
\end{lemma}
The proof is obvious.
\begin{lemma}
  If $\langle S,f\rangle$ is a case of finitary funny business, then
  there are $A, B \subset S$ and their two elementary possibilities,
  resp., $\Pi_A\langle h_A\rangle$, $\Pi_B\langle h_B\rangle$ which
  constitute a case of generalized primary slr modal-correlation funny
  business.
\end{lemma}

\noindent
{Proof:} From the definition of finitary funny business, we have $A$ and $B$
such that $A \; SLR\; B$ and $\bigcap \{ f(e) : e \in A \} \neq
\emptyset$, $\bigcap \{ f(e) : e \in B \} \neq \emptyset$. For $h_A$
take then any history from the first intersection, and for $h_B$-- any
history from the other. Hence $A \subset h_A$ and $B \subset h_B$.
From $\bigcap \{ f(e) : e \in A \cup B \} = \emptyset$ it follows that
$\Pi_A\langle h_A\rangle \cap \Pi_B\langle h_B\rangle =
\emptyset$. \quad $\square$

Turning next to combinatorial funny business, \citet{muller06:_funny}
define it in terms of a set $T$ of elementary transition, where
elementary transition is a pair: $\langle$ point event $e_i$,
elementary possibility $H_i$ at $e_i \; \rangle$. Thus, a given
$\langle S, f \rangle$ uniquely specifies a set of of transitions, and
a set of transition uniquely determines a pair $\langle S, f\rangle$.
To ease the exposition, in the lemmas below we will refer to $\langle
S, f \rangle$ as a set of transitions, and we will claim that under
certain condition, $\langle S, f \rangle$ is a case of infinitary
funny business iff $\langle S, f \rangle$ is a case of combinatorial
funny business.

\citet{muller06:_funny} say that set $T$ of elementary transitions is
{\em combinatorially consistent} iff for any two transitions $t_i, t_j
\in T$, (1) if $e_i = e_j$, then $H_i = H_j$, (2) if $e_i < e_j$, then
$H_{(e_j)} \subseteq H_i$, (3) if $e_j < e_i$, then $H_{(e_i)}
\subseteq H_j$, and (4) if $e_i$ and $e_j$ are incomparable, then $e_i
\;SLR \; e_j$.

Finally, they say that $T$ is consistent iff $\emptyset \neq H_T :=
\bigcap \{H_i \mid H_i \in t_i \wedge t_i \in T\}$ and that $T$
constitutes a case of combinatorial funny business iff $T$ is
combinatorially consistent but inconsistent.

We finally turn to our  lemmas:

\begin{lemma}
  If $\langle S, f\rangle$ is a case of infinitary funny business,
  then $\langle S, f\rangle$ constitutes a case of combinatorial funny
  business.
\end{lemma}

\noindent {Proof: } We need first to show that $\langle S, f\rangle$
is consistent. Since $f$ is a function, condition (1) of combinatorial
consistency holds.  Conditions (2) and (3) of combinatorial
consistency follow from the condition: $\forall e, e' \in S\; (e < e'
\rightarrow f(e') \subseteq f(e))$ of infinitary funny business.
(Suppose $e < e'$. Let $f(e') = H_j$ and $f(e) = H_i$. Since $f(e')
\subseteq f(e)$, then $H_j \subseteq H_i$. We need to prove that
$H_{(e')} \subseteq H_i$. Suppose to the contrary that (a) there is a
history $h^* \in H_{(e')} - H_i$. Since $e < e'$, there exists a
history $h \in H_{e'} \cap H_i$. From the fact that $e' \in h \cap
h^*$ we get that $h \equiv_e h^*$, therefore $h^* \in H_i$ and we
arrive at a contradiction with (a).) Condition (4) of combinatorial
consistency comes from the condition $\forall A\; (A \subset_{fin} S
\rightarrow \exists\! h \in Hist: \; A \subset h)$ of infinitary funny
business. Finally, since $H_T = \bigcap \{ f(e) : e \in S \}$, by
$\bigcap \{ f(e) : e \in S \} = \emptyset$ we get $H_T
=\emptyset$.\quad $\square$

\begin{lemma}
  If $\langle S, f\rangle$ constitutes a case of combinatorial funny
  business but $\langle S, f\rangle$ is not a case of finitary funny
  business, then $\langle S, f\rangle$ is a case of infinitary funny
  business.
\end{lemma}

\noindent
{Proof: } From $H_T = \emptyset$ and $H_T = \bigcap \{ f(e) : e \in S
\}$, it must be that $\bigcap \{ f(e) : e \in S \} = \emptyset$.
Conditions (2) and (3) of consistency entail condition: $\forall e, e'
\in S\; (e < e' \rightarrow f(e') \subseteq f(e))$ of infinitary funny
business.  And, if $card(S) < \omega$, by condition (4) of
combinatorial consistency and the assumption that $\langle S,
f\rangle$ is not a case of finitary funny business, we get that
$\langle S, f\rangle$ is consistent, i.e., not a case of combinatorial
funny business. Thus, $card(S) \geqslant \omega$. \quad $\square$

\section{Location of choice points}

Consider Prior Choice Principle. It says that if (0) $x\in h/h'$, then
there is $e$ such that (1) $e < x$ and (2) $h' \perp_e h$. (0) - (2)
then entail that $\forall h\; x \in h \rightarrow h' \perp_e h$, which
can be written as $h' \perp_e H_{(x)}$. Thus, defining $C_{h'}(x) :=
\{e \in W\mid h' \perp_e H_{(x)}\}$, we already know that if
$C_{h'}(x)$ is non-empty, then by (1) at least one of its elements is
below $x$. Where are other elements of $C_{h'}(x)$ located? Our claim
now is that if (3) $C_{h'}(x) \cup \{x\}$ does not give rise to FINFB,
then $C_{h'}(x) < x $. For reductio, suppose that $h' \perp_{e'}
H_{(x)}\}$ and $e' \; SLR\; x$. By the assumption (3), there is $h''
\in \Pi_{e'}\langle h' \rangle \cap H{(x)}$, and hence $h'
\equiv_{e'} h''$ and $h'' \in H_{(x)}$, which contradicts $h'
\perp_{e'} H_{(x)}$.

For a future reference, we will generalize slightly these observations
and put it as a fact:

\begin{fact}
  If $C(x) := \{e \in W\mid \exists \! h: h \perp_e H_{(x)}\}$ is
  non-empty and $C(x) \cup \{x\}$ does not give rise to FINFB, then
  $C(x) < x $.
\label{nbtheorem}
\end{fact}

\noindent
It is worth noting that the fact is a special case of one of
equivalences of four various notions of modal funny business that
\citet{belnap_nato} and \citet{belnap03_no_cc} established. It is a
special case of the implication from the so-called {\em
  some-cause-like-locus-not-in-past funny business} to generalized
primary slr modal-correlation funny business.


\end{document}